Preprint September 2026

# Sequential Bayesian diagnosis and prognosis of structural deterioration from measured structural responses

Ronald Schneider

Bundesanstalt für Materialforschung und -prüfung (BAM), Unter den Eichen 87, 12205 Berlin, Germany

Email: ronald.schneider@bam.de

## Abstract

Structural health monitoring of deteriorating structures requires the latent deterioration process to be inferred indirectly from measured structural responses and propagated to quantities of engineering interest. This paper presents a probabilistic framework that formalizes this process by distinguishing Bayesian inference of uncertain deterioration and model parameters from diagnosis of the current deterioration state, prognosis of its future evolution, and prediction of structural reliability. The deterioration model thereby provides the common probabilistic link between monitoring-based inference and these downstream tasks. To enable sequential inference as monitoring information accumulates, Bayesian Updating with Structural Reliability Methods (BUS) is extended through a nested formulation of the BUS events associated with successively growing monitoring datasets. The resulting Sequential BUS method allows Subset Simulation to continue from the augmented conditional sample population obtained at the preceding monitoring stage rather than restarting from the prior distribution. The posterior population can then be propagated to monitoring-informed reliability analysis. The framework is demonstrated for fatigue deterioration of a redundant jacket-type structure, where modal properties obtained from simulated vibration monitoring are used to infer a high-dimensional deterioration model of welded connections. The results show how successive monitoring observations constrain plausible deterioration histories, provide probabilistic system-level diagnosis, and update crack-growth prognosis and structural reliability while retaining ambiguity between deterioration scenarios that produce similar structural responses. The proposed framework thus provides a coherent probabilistic route from response-based monitoring data to deterioration diagnosis, prognosis, and reliability prediction.

## Keywords

Structural health monitoring; Sequential Bayesian inference; Deterioration diagnosis and prognosis; Bayesian Updating with Structural Reliability Methods (BUS); Structural reliability

# 1 Introduction

Structural health monitoring (SHM) of deteriorating structures seeks to estimate structural condition from measured structural responses. In the classical SHM setting considered here, the deterioration state is latent: monitoring systems observe its effects on structural behavior rather than the deterioration process itself. SHM is therefore naturally formulated as an inverse problem in which response measurements, or features derived from them such as modal properties, are used to estimate quantities describing structural condition. The associated tasks are commonly classified as damage detection, localization, quantification, and prognosis [1-3]. Detection may be performed directly from changes in response features (e.g., [4]), whereas localization and quantification generally require a model that maps structural condition to observable behavior (e.g., [5, 6]). Prognosis further requires a model for the temporal evolution of that condition.

The present work focuses on the model-based setting advocated for structural and infrastructure systems [7], in which monitoring information is interpreted through models of structural condition, behavior, and performance to support condition assessment, prediction of future performance, and ultimately monitoring-informed decision-making. A probabilistic formulation requires a deterioration model describing the physical condition and its evolution, a structural behavior model relating that condition to observable quantities, and a prediction-error model relating predicted and measured behavior. Together, these models define the probabilistic relation between the latent deterioration process and the monitoring observations required to infer uncertain deterioration and model parameters from structural response data. If structural safety is of interest, the monitoring-informed deterioration uncertainty must additionally be propagated through a performance model relating structural condition to performance or failure. The behavior model therefore provides the observation link required for inference, whereas the performance model provides the consequence link required for reliability prediction [8].

Previous studies have developed the main elements of this framework through the extension of inspection-informed reliability updating to response-based monitoring. Within structural reliability, Bayesian updating of deterioration and reliability from inspection information has a long history across different deterioration mechanisms. Among these developments, a prominent line of work concerned fatigue of welded connections [9-12], which is also the deterioration mechanism considered in the numerical application of this paper. In these approaches, relatively direct observations of component condition are used to infer uncertain deterioration-model parameters and update component reliability. Such approaches were subsequently extended to deteriorating structural systems, where inspection-informed deterioration uncertainty is propagated to system reliability [13, 14].

In parallel, Bayesian model updating developed the response-based inference capabilities required for SHM. Early developments focused on system identification, using structural behavior models and measured responses to infer uncertain structural-model parameters and their posterior uncertainty [15-17]. These concepts were subsequently extended to damage identification by interpreting changes in damage-sensitive structural parameters, particularly stiffness parameters, in terms of the probability, location, and severity of structural damage [18-20].

These developments provided complementary ingredients. Structural reliability research contributed explicit deterioration models and methods for propagating deterioration uncertainty to structural system reliability, while Bayesian model updating provided structural behavior models and inference methods for extracting information on structural condition indirectly from measured responses. Their integration with explicit temporal deterioration models enabled the parameters governing deterioration evolution, and hence current and future deterioration states, to be inferred from response-monitoring data and the resulting uncertainty to be propagated to structural reliability. Kamariotis et al. [21] demonstrated this using vibration-based modal data and gradual deterioration models, with the inferred deterioration uncertainty propagated to structural reliability and life-cycle decisions. Their subsequent framework [22] extended the approach to stochastic gradual and shock deterioration and sequential joint state-parameter estimation using monitoring and inspection information, again linking deterioration inference to reliability updating and life-cycle decision analysis.

These developments demonstrate the feasibility of combining response-based monitoring, deterioration modelling, and structural reliability analysis. However, the role of the deterioration model as the central link between monitoring-based inference and subsequent SHM and engineering tasks has remained largely implicit. In a model-based probabilistic formulation, monitoring data do not directly provide diagnosis, prognosis, structural behavior predictions, or reliability estimates. Rather, the data are used, through the structural behavior and prediction-error models, to infer the uncertain deterioration process and its governing parameters. The resulting posterior deterioration model provides the common basis for subsequent tasks: its current state characterizes deterioration diagnosis, its propagation in time provides prognosis, and its further propagation through structural behavior or performance models yields monitoring-informed response and reliability predictions. Distinguishing these stages clarifies what is inferred from monitoring data, what is obtained through posterior prediction, and which models are required for each task. Although these elements have been combined in previous applications, their roles and interrelations within a general probabilistic architecture for response-based SHM have not been formulated systematically.

The first objective of this paper is therefore to formalize such an architecture for response-based SHM of deteriorating structures. The framework places Bayesian inference, deterioration diagnosis and prognosis, structural behavior prediction, and structural reliability prediction within a common probabilistic formulation while explicitly distinguishing inference from posterior prediction. Both batch and sequential Bayesian inference are considered as monitoring information becomes available over time. The contribution lies not in introducing these individual tasks, but in systematically defining their relationships within a common framework.

The second objective addresses the computational implementation of sequential Bayesian inference within this framework. As new response data become available, the posterior deterioration model must be updated repeatedly. Existing approaches to sequential deterioration inference include recursive Bayesian filtering and repeated batch inference [23]. Here, Bayesian Updating with Structural Reliability Methods (BUS) [24] is extended to accommodate successively growing monitoring datasets. BUS introduces an auxiliary random variable and defines an event through the likelihood such that conditioning on this event yields the Bayesian posterior distribution. Interpreting this event as a failure event enables simulation-based structural reliability methods to generate posterior samples and estimate the Bayesian model evidence; Subset Simulation [25] is commonly used for this purpose. Under sequential monitoring, the updating problems associated with successively growing datasets are represented by a nested sequence of BUS events. The conditional sample population obtained at one monitoring stage can therefore be further conditioned as new data become available, rather than restarting the inference from the original prior distribution. The resulting Sequential BUS method exploits the information accumulated during preceding monitoring stages.

A distinct advantage of this construction is that the posterior samples generated by the sequential inference can subsequently serve as the starting population for monitoring-informed reliability analysis. A two-phase BUS/Subset Simulation strategy has been described in which samples conditional on a Bayesian updating event are carried forward to posterior rare-event analysis [26]. This strategy has subsequently been applied to deteriorating structural systems, with samples obtained from monitoring- or inspection-informed updating used directly to initialize Subset Simulation for estimating conditional system failure probabilities [14, 27]. Related two-stage formulations have also been used to initialize updated first-passage reliability analysis from posterior samples obtained by Bayesian system identification [28]. The present formulation extends this computational chain to sequential monitoring: sample populations are reused first between successive Bayesian updating stages and subsequently from the updated posterior in rare-event reliability analysis. Sequential BUS with Subset Simulation thereby provides an integrated computational implementation of sequential deterioration inference and monitoring-informed structural reliability prediction.

The proposed probabilistic framework and Sequential BUS method are demonstrated for fatigue deterioration of a redundant jacket-type structure. Modal properties obtained from simulated vibration responses are used sequentially to infer a high-dimensional fatigue deterioration model for the welded connections. The example considers system-level deterioration diagnosis, hotspot deterioration prognosis, and monitoring-informed structural reliability prediction, thereby illustrating both the probabilistic framework and its Sequential BUS implementation.

The remainder of the paper is organized as follows. Section 2 develops the general probabilistic framework, including the deterioration, structural behavior, prediction-error, and performance models, and distinguishes Bayesian inference from monitoring-informed diagnosis, prognosis, structural behavior prediction, and reliability prediction. Section 3 presents the Sequential BUS formulation and its implementation with Subset Simulation for sequential Bayesian inference. Section 4 demonstrates the framework and computational method for fatigue deterioration of a redundant jacket-type structure using modal monitoring information. Section 5 discusses model requirements, computational efficiency and robustness of Sequential BUS and its Subset Simulation implementation, identifiability and monitoring-system design considerations, and the engineering implications and scope of the proposed framework. Section 6 concludes the paper.

# 2 General probabilistic framework

## 2.1 Probabilistic formulation of the monitoring model

The proposed framework relates structural monitoring data to the deterioration state of a structural system through a deterioration model, a structural behavior model, and a prediction-error model. Together, these models establish the probabilistic relationship between the uncertain model parameters and the monitoring observations and thereby define the likelihood required for Bayesian inference in Section 2.2. For clarity, the following formulation adopts deterministic deterioration and structural behavior models conditional on their uncertain parameters. Loads or other inputs required by the structural behavior models are assumed to be known and deterministic. The formulation considers static inputs and corresponding structural response quantities evaluated at discrete monitoring stages, as well as structural properties such as modal quantities that do not require an explicit input. Dynamic excitation and time-history response models are not considered explicitly. These assumptions are made for simplicity and are not fundamental limitations of the general framework. The framework further presupposes that suitable deterioration and structural behavior models are available and that the monitoring system provides measurements that are informative with respect to deterioration mechanisms relevant to structural performance and lifetime-management decisions.

**Deterioration model:** The formulation of the deterioration model starts with identifying the deterioration processes and structural components that are relevant to the monitoring problem. In particular, deterioration mechanisms and components should be considered if their deterioration may affect structural behavior, structural performance, or ultimately lifetime-management decisions. This requirements analysis determines which deterioration processes and components need to be represented and the level of detail required.

Deterioration models are typically formulated at the structural-component level. The framework therefore requires an appropriate set of deterioration models and associated uncertain parameters to describe the time-dependent evolution of deterioration in the relevant components of the physical structure. Let $\boldsymbol{D}(t)$ denote the component-resolved deterioration state of the structural system at time $t$. Its evolution is represented by the vector-valued deterioration model

$$\boldsymbol{D}(t) = \boldsymbol{h}_D(t; \boldsymbol{\Theta}_D) \tag{1}$$

where $\boldsymbol{h}_D = [h_{D,1}, \dots, h_{D,n}]^T$ collects the deterioration models of the relevant structural components, and $\boldsymbol{\Theta}_D$ contains the associated uncertain deterioration-model parameters, including the initial deterioration states where these are uncertain. The system deterioration state is obtained by jointly evaluating the individual component deterioration models. Dependencies between component deterioration states can be represented through dependencies among the corresponding deterioration-model parameters.

The individual entries of $\boldsymbol{D}(t)$ should represent physically meaningful deterioration quantities, such as fatigue-crack size, corrosion loss, number of broken wires, or other component-specific measures of deterioration. This is important because the deterioration state provides the physical link between the deterioration processes and their effects on structural behavior. The deterioration models should therefore represent the physically plausible and decision-relevant deterioration states that the monitoring system is intended to diagnose and prognose.

For a realization $\boldsymbol{\theta}_D$ of the deterioration-model parameters,

$$\boldsymbol{d}(t) = \boldsymbol{h}_D(t; \boldsymbol{\theta}_D) \tag{2}$$

denotes the corresponding deterministic evolution of the system deterioration state.

**Structural behavior model:** The structural behavior model links the deterioration state of the system to quantities that can be observed through monitoring. Depending on the monitoring system, one or several behavior models may be required to represent different aspects of the static or dynamic structural response, for example modal frequencies, mode shapes, strains, or displacements. Such behavior models are commonly finite element models and may possess model-specific parameters as well as parameters shared across several models.

At monitoring stage $k$, the predicted structural behavior is represented generically by

$$\boldsymbol{Y}_k = \boldsymbol{h}_B(\boldsymbol{D}(t_k), \boldsymbol{u}_k; \boldsymbol{\Theta}_B) \tag{3}$$

where $\boldsymbol{h}_B$ denotes the structural behavior mapping obtained from one or several behavior models, each of which may predict multiple observable quantities, and $\boldsymbol{\Theta}_B$ contains the associated uncertain model parameters, and $\mathbf{u}_k$ denotes known deterministic inputs associated with monitoring stage $k$. For static response measurements, $\mathbf{u}_k$ may, for example, represent the applied static loads. For quantities such as modal properties obtained from eigenvalue analysis, no explicit input $\mathbf{u}_k$ is required.

The type, level of detail, and parameterization of the behavior models should be selected such that, conditional on the relevant deterioration states, they adequately predict the quantities observed by the monitoring system. In particular, the measured structural behavior must be sufficiently sensitive to the deterioration mechanisms and components represented by the deterioration model for the monitoring data to provide information about deterioration.

For given realizations $\boldsymbol{\theta}_D$ and $\boldsymbol{\theta}_B$, the predicted behavior at monitoring stage $k$ is

$$\boldsymbol{y}_k(\boldsymbol{\theta}_D, \boldsymbol{\theta}_B; \boldsymbol{u}_k) = \boldsymbol{h}_B(\boldsymbol{h}_D(t_k; \boldsymbol{\theta}_D), \boldsymbol{u}_k; \boldsymbol{\theta}_B) \tag{4}$$

This composition explicitly links the deterioration state predicted by the deterioration model to the structural quantities that can be compared with monitoring observations.

**Prediction-error model:** The behavior-model predictions are related to the monitoring observations through a probabilistic prediction-error model. Let $\boldsymbol{Z}_k$ denote the random vector of monitoring observations obtained at monitoring stage $k$. The prediction-error model describes the discrepancy between the observations and the corresponding behavior-model predictions and may account for measurement uncertainty and model uncertainty. It is characterized by uncertain parameters $\boldsymbol{\Theta}_{\mathcal{E}}$.

One possible formulation is an additive prediction-error model,

$$\boldsymbol{Z}_k = \boldsymbol{y}_k(\boldsymbol{\Theta}_D, \boldsymbol{\Theta}_B; \boldsymbol{u}_k) + \boldsymbol{\mathcal{E}}_k, \tag{5}$$

or equivalently,

$$\boldsymbol{\mathcal{E}}_k = \boldsymbol{Z}_k - \boldsymbol{y}_k(\boldsymbol{\Theta}_D, \boldsymbol{\Theta}_B; \boldsymbol{u}_k) \tag{6}$$

where $\mathcal{E}_k$ denotes the random prediction error at monitoring stage $k$. The additive formulation is adopted here for simplicity but is not a restriction of the general framework. Other probabilistic formulations relating the behavior-model predictions to the monitoring observations may be used where appropriate. The prediction-error model used here combines measurement uncertainty and model uncertainty in a single probabilistic term. More detailed formulations may represent these uncertainty sources separately [29, 30], which leads to alternative likelihood formulations. The combined formulation is adopted here for simplicity and to keep the subsequent development concise.

For monitoring data collected up to stage $k$, define $\boldsymbol{Z}_{1:k} = [\boldsymbol{Z}_1^T, \dots, \boldsymbol{Z}_k^T]^T$ and $\boldsymbol{y}_{1:k}(\boldsymbol{\Theta}_D, \boldsymbol{\Theta}_B; \boldsymbol{u}_{1:k}) = [\boldsymbol{Y}_1^T, \dots, \boldsymbol{Y}_k^T]^T$. Under the additive formulation, the corresponding prediction-error history is

$$\boldsymbol{\mathcal{E}}_{1:k} = \boldsymbol{Z}_{1:k} - \boldsymbol{y}_{1:k}(\boldsymbol{\Theta}_D, \boldsymbol{\Theta}_B; \boldsymbol{u}_{1:k}). \tag{7}$$

Its joint probability density is denoted by

$$f_{\boldsymbol{\mathcal{E}}_{1:k}}(\boldsymbol{\varepsilon}_{1:k} | \boldsymbol{\theta}_{\mathcal{E}}). \tag{8}$$

where $\boldsymbol{\theta}_{\mathcal{E}}$ denotes a realization of the prediction-error-model parameters $\boldsymbol{\Theta}_{\mathcal{E}}$. The joint formulation does not require prediction errors at different monitoring times or for different measured quantities to be statistically independent. Temporal, spatial, or other dependencies may therefore be represented through the prediction-error model.

Collecting the uncertain parameters of the deterioration, behavior, and prediction-error models gives the complete parameter vector

$$\boldsymbol{\Theta} = [\boldsymbol{\Theta}_D^T, \boldsymbol{\Theta}_B^T, \boldsymbol{\Theta}_{\mathcal{E}}^T]^T \tag{9}$$

with realization

$$\boldsymbol{\theta} = [\boldsymbol{\theta}_D^T, \boldsymbol{\theta}_B^T, \boldsymbol{\theta}_{\mathcal{E}}^T]^T \tag{10}$$

The deterioration-model parameters $\boldsymbol{\Theta}_D$ are of primary interest for monitoring-based inference and updating, whereas $\boldsymbol{\Theta}_B$ and $\boldsymbol{\Theta}_{\mathcal{E}}$ may be treated as additional uncertain parameters in the Bayesian inference.

For realized monitoring observations $\mathbf{z}_{1:k}$ and given parameter realizations $\boldsymbol{\theta}_D$ and $\boldsymbol{\theta}_B$, the corresponding prediction-error realization is

$$\boldsymbol{\varepsilon}_{1:k}(\boldsymbol{\theta}_D, \boldsymbol{\theta}_B) = \boldsymbol{z}_{1:k} - \boldsymbol{y}_{1:k}(\boldsymbol{\theta}_D, \boldsymbol{\theta}_B; \boldsymbol{u}_{1:k}). \tag{11}$$

The likelihood of the observed monitoring data is therefore

$$L_{1:k}(\boldsymbol{\theta}) = f_{\boldsymbol{\mathcal{E}}_{1:k}}(\boldsymbol{\varepsilon}_{1:k}(\boldsymbol{\theta}_D, \boldsymbol{\theta}_B)|\boldsymbol{\theta}_{\mathcal{E}}) \tag{12}$$

Because the inputs $\boldsymbol{u}_{1:k}$ are treated as known, their dependence is suppressed in the likelihood notation $L_{1:k}(\boldsymbol{\theta})$ in the following. This likelihood forms the basis for the Bayesian inference described in Section 2.2.

## 2.2 Bayesian inference from monitoring data

Section 2.1 describes the probabilistic mapping from the uncertain model parameters to the monitoring observations. Bayesian inference reverses this relationship probabilistically: observed monitoring data are used to update the uncertainty in the model parameters. Before monitoring data are assimilated, the uncertain parameter vector $\boldsymbol{\Theta}$ is described by the prior density $p(\boldsymbol{\theta})$.

The prior may incorporate information available before deterioration monitoring begins. In particular, measurements of the structural behavior in the initial or nominal structural state may be used for system identification and calibration of the structural behavior and prediction-error models. The resulting knowledge of $\boldsymbol{\Theta}_B$ and $\boldsymbol{\Theta}_{\mathcal{E}}$ can then be represented in the prior and propagated in the subsequent inference of the deterioration-model parameters $\boldsymbol{\Theta}_D$.

Given monitoring observations $\mathbf{z}_{1:k}$ collected up to monitoring stage $k$, Bayes' rule gives

$$p(\boldsymbol{\theta}|\boldsymbol{z}_{1:k}) = \frac{L_{1:k}(\boldsymbol{\theta})\, p(\boldsymbol{\theta})}{c_{E,1:k}} \tag{13}$$

where $L_{1:k}(\boldsymbol{\theta})$ is the likelihood defined in Section 2.1 and

$$c_{E,1:k} = p(\boldsymbol{z}_{1:k}) = \int L_{1:k}(\boldsymbol{\theta})\, p(\boldsymbol{\theta})\, d\boldsymbol{\theta} \tag{14}$$

is the evidence. Equation (13) represents batch inference, in which all monitoring data available up to stage $k$are treated jointly.

In structural health monitoring, however, observations typically become available sequentially. The contribution of the new observations $\mathbf{z}_k$, conditional on the previously available observations, is described by the conditional likelihood

$$L_{k|1:k-1}(\boldsymbol{\theta}) = p(\boldsymbol{z}_k|\boldsymbol{z}_{1:k-1}, \boldsymbol{\theta}). \tag{15}$$

For the additive prediction-error formulation introduced in Section 2.1, the conditional likelihood can be obtained directly from the joint prediction-error distribution as

$$L_{k|1:k-1}(\boldsymbol{\theta}) = \frac{f_{\mathcal{E}_{1:k}}(\boldsymbol{\varepsilon}_{1:k}(\boldsymbol{\theta}_D, \boldsymbol{\theta}_B)|\boldsymbol{\theta}_{\mathcal{E}})}{f_{\mathcal{E}_{1:k-1}}(\boldsymbol{\varepsilon}_{1:k-1}(\boldsymbol{\theta}_D, \boldsymbol{\theta}_B)|\boldsymbol{\theta}_{\mathcal{E}})} \tag{16}$$

Hence, the conditional likelihood can be constructed from the joint prediction-error model without assuming conditional independence between monitoring stages. Any temporal, spatial, or other dependence represented by the joint prediction-error distribution is thereby retained in the sequential Bayesian inference.

By the probability chain rule, the cumulative likelihood satisfies

$$L_{1:k}(\boldsymbol{\theta}) = L_{1:k-1}(\boldsymbol{\theta})\, L_{k|1:k-1}(\boldsymbol{\theta}). \tag{17}$$

Bayes' rule can therefore be written recursively as

$$p(\boldsymbol{\theta}|\mathbf{z}_{1:k}) = \frac{L_{k|1:k-1}(\boldsymbol{\theta})\, p(\boldsymbol{\theta}|\boldsymbol{z}_{1:k-1})}{c_{E,k|1:k-1}}, \tag{18}$$

where

$$c_{E,k|1:k-1} = p(\boldsymbol{z}_k|\boldsymbol{z}_{1:k-1}) = \int L_{k|1:k-1}(\boldsymbol{\theta})\, p(\boldsymbol{\theta}|\boldsymbol{z}_{1:k-1})\, d\boldsymbol{\theta} \tag{19}$$

is the conditional evidence associated with the newly available monitoring data. The cumulative and conditional evidences are related by

$$c_{E,1:k} = c_{E,1:k-1}\, c_{E,k|1:k-1} \tag{20}$$

Equations (13) and (18) define, respectively, the batch and sequential Bayesian inference problems arising in structural health monitoring. The former infers the model parameters jointly from all available monitoring data, whereas the latter updates the posterior recursively as new monitoring information becomes available. For complex structural models and informative monitoring data, direct evaluation of these posterior distributions can be computationally challenging. Section 3 therefore formulates their solution using BUS [24].

### 2.3 Monitoring-informed engineering predictions

The Bayesian inference of Section 2.2 provides the posterior distribution $p(\boldsymbol{\theta}|\mathbf{z}_{1:k})$ of the uncertain deterioration, behavior, and prediction-error model parameters given the monitoring data available up to stage $k$. In the proposed framework, this inference is distinguished from the subsequent engineering tasks performed with the updated probabilistic model. The posterior uncertainty in the deterioration-model parameters is propagated through the deterioration model to characterize current and future deterioration states These monitoring-informed deterioration states form the basis for diagnosis and prognosis and can subsequently be propagated through structural models for structural response and reliability prediction. Thus, Bayesian updating itself does not constitute diagnosis, prognosis, structural behavior prediction, or reliability prediction; rather, these are subsequent posterior-propagation tasks based on the inferred model uncertainty.

**Diagnosis:** Diagnosis concerns the deterioration state of the structure at the current monitoring time $t_k$. From the deterioration model, $\boldsymbol{D}(t_k) = \boldsymbol{h}_D(t_k; \boldsymbol{\Theta}_D)$, and propagation of the posterior uncertainty in $\boldsymbol{\Theta}_D$ gives the monitoring-informed distribution

$$p(\boldsymbol{d}(t_k)|\boldsymbol{z}_{1:k}). \tag{21}$$

In the terminology adopted here, diagnosis encompasses the classical SHM damage-identification tasks of detection, localization, and quantification. Damage detection concerns whether relevant deterioration is present in the system, localization concerns which structural components are affected, and quantification concerns the extent of deterioration in these components.

Let $d_{j,\lim}$ denote a threshold defining a relevant deterioration state for component $j$. The posterior probability that deterioration exceeding these thresholds has occurred somewhere in the system can be expressed as

$$p_{\text{dam}}(t_k|\boldsymbol{z}_{1:k}) = 1 - \Pr[D_j(t_k) \leq d_{j,\lim}\ \forall j \mid \boldsymbol{z}_{1:k}] \tag{22}$$

This probability provides a basis for damage detection. Conceptually, a nonzero value of $p_{\mathrm{dam}}(t_k|\mathbf{z}_{1:k})$ indicates posterior probability that relevant deterioration is present somewhere in the system. In practice, however, uncertainty in the inferred deterioration state will generally result in a nonzero posterior probability of deterioration even when the monitoring data provide little indication of damage. An operational detection decision can therefore be defined by

$$p_{\mathrm{dam}}(t_k|\boldsymbol{z}_{1:k}) > p_{\mathrm{th}} \tag{23}$$

where $p_{\mathrm{th}}$ a prescribed detection threshold.

Localization can be characterized through the component-specific probabilities

$$p_{\mathrm{dam},j}(t_k|\boldsymbol{z}_{1:k}) > \Pr[D_j(t_k) > d_{j,\mathrm{lim}} \mid \boldsymbol{z}_{1:k}], \tag{24}$$

whereas quantification is provided by the posterior distributions

$$p(d_j(t_k)|\boldsymbol{z}_{1:k}) \tag{25}$$

of the individual component deterioration states.

**Prognosis:** Prognosis concerns the future evolution of deterioration. For a future time $t > t_k$, propagation of the posterior uncertainty in the deterioration-model parameters gives

$$p(\boldsymbol{d}(t)|\boldsymbol{z}_{1:k}), \quad t > t_k, \tag{26}$$

which represents the monitoring-informed prognosis of the system deterioration state. Diagnosis and prognosis therefore follow from the same posterior deterioration model and differ only in the time at which the deterioration state is evaluated: diagnosis concerns the current state, whereas prognosis concerns its future evolution.

**Structural behavior prediction:** The posterior distribution obtained from monitoring can also be used to predict structural behavior at the current or a future evaluation stage. For a prescribed deterministic input $\mathbf{u}_r$, the behavior-model prediction at time $t_r$ is $\boldsymbol{Y}_r = \boldsymbol{h}_B(\boldsymbol{D}(t_r), \boldsymbol{u}_r, \boldsymbol{\Theta}_B)$, gives the posterior distribution of the behavior-model prediction

$$p(\boldsymbol{y}_r|\boldsymbol{z}_{1:k}, \boldsymbol{u}_r). \tag{27}$$

For static response quantities, $\boldsymbol{u}_r$ represents the prescribed static load or other known input under which the structural behavior is predicted. For quantities such as modal properties, no explicit input is required. Consistent with the formulation in Section 2.1, dynamic excitation and time-history response predictions are not considered explicitly.

The distribution $p(\boldsymbol{y}_r|\boldsymbol{z}_{1:k}, \boldsymbol{u}_r)$ represents the uncertainty in the behavior-model prediction resulting from the posterior uncertainty in the deterioration and behavior-model parameters. If the prediction-error model additionally represents uncertainty associated with imperfections of the structural behavior model, this model uncertainty should also be included when predicting the actual structural behavior. Measurement uncertainty, in contrast, is required only when predicting future monitoring observations.

The structural behavior model therefore has two related roles in the framework. First, it links deterioration to the observed structural behavior for Bayesian inference. Second, after updating, it provides the model mapping used for structural behavior prediction, with any additional model uncertainty included according to the adopted prediction-error formulation.

**Structural reliability prediction:** Structural reliability prediction represents a further downstream use of the monitoring-informed posterior distribution. The structural model required for reliability assessment does not need to be identical to the behavior model used for monitoring-based inference (see also [8]). For example, monitoring inference may rely on a structural dynamics model, whereas reliability assessment may require a separate nonlinear static finite element model for evaluating structural capacity. A structural performance model can therefore be introduced generically as

$$R(t) = h_P(\boldsymbol{D}(t); \boldsymbol{\Theta}_B, \boldsymbol{\Theta}_P) \tag{28}$$

where $h_P(\cdot)$ maps the deterioration state $\boldsymbol{D}(t)$ and uncertain model parameters to the time-dependent structural capacity $R(t)$, which may be evaluated for a prescribed loading configuration. The parameter vector $\boldsymbol{\Theta}_P$ contains parameters specific to, or additionally required by, the performance model. Parameters in $\boldsymbol{\Theta}_B$ and $\boldsymbol{\Theta}_P$ that are shared with, or statistically related to, quantities entering the behavior model may be informed by the monitoring data through the Bayesian inference described in Section 2.2.

For the resistance–load formulation considered here, let $S(t)$ denote the probabilistic load or load effect. Structural failure is defined through the time-dependent limit-state function

$$g(t) = R(t) - S(t), \tag{29}$$

and the monitoring-informed cumulative failure probability up to time $t$ is

$$p_F(t|\boldsymbol{z}_{1:k}) = \Pr\left(\min_{\tau\in[0,t]} g(\tau) \leq 0 \mid \boldsymbol{z}_{1:k}\right) \tag{30}$$

The reliability prediction is thus obtained by propagating the monitoring-informed posterior uncertainty through the deterioration and performance models and combining the resulting uncertainty in structural capacity with the probabilistic load model (e.g., [31]). In contrast to the behavior model, which provides the observation link required for inference, the performance model provides the consequence link through which the inferred deterioration and parameter uncertainty affects structural reliability.

# 3 Bayesian updating with structural reliability methods

Section 2.2 formulated two Bayesian inference problems arising in structural health monitoring of deteriorating structural systems: batch inference based on all monitoring data available up to a given stage, and sequential inference in which the posterior distribution is updated recursively as new monitoring data become available. This section first reviews BUS and its implementation with Subset Simulation for the batch inference problem. In BUS, Bayesian updating is reformulated as conditional sampling in an augmented probability space by defining an event through the likelihood, such that samples conditional on this event follow the posterior distribution.

Building on the classical BUS formulation, the main methodological development of this section is a sequential extension for recursive Bayesian updating. The key idea is to construct the BUS events associated with successive monitoring stages as nested events, such that the augmented conditional sample population obtained at one stage can be carried forward to the next update. This nested-event construction is subsequently combined with Subset Simulation, allowing the conditional sampling sequence to be continued across monitoring stages rather than restarted from the augmented prior distribution.

## 3.1 BUS

The batch Bayesian inference problem defined in Section 2.2 is first reformulated using BUS. BUS augments the uncertain parameter vector $\boldsymbol{\Theta}$ by an independent auxiliary random variable $\Pi \sim \mathcal{U}(0,1)$. A positive likelihood multiplier $c_{1:k}$ is selected such that $c_{1:k}\, L_{1:k}(\boldsymbol{\theta}) \leq 1$ for all $\boldsymbol{\theta}$, which requires $c_{1:k} \leq [\sup_{\boldsymbol{\theta}} L_{1:k}(\boldsymbol{\theta})]^{-1}$.

The BUS event is defined as

$$\mathcal{O}_{1:k} = \{\Pi \leq c_{1:k}\, L_{1:k}(\boldsymbol{\Theta})\}. \tag{31}$$

For a specified parameter realization $\boldsymbol{\Theta} = \boldsymbol{\theta}$, the probability of the BUS event is $\Pr(\mathcal{O}_{1:k} | \boldsymbol{\Theta} = \boldsymbol{\theta}) = \Pr[\Pi \leq c_{1:k}\, L_{1:k}(\boldsymbol{\theta})]$. Since $\Pi \sim \mathcal{U}(0,1)$,

$$\begin{aligned} \Pr(\mathcal{O}_{1:k} | \boldsymbol{\Theta} = \boldsymbol{\theta}) &= \int_0^{c_{1:k}\, L_{1:k}(\boldsymbol{\theta})} p(\pi)\, du \\ &= \int_0^{c_{1:k}\, L_{1:k}(\boldsymbol{\theta})} 1\, du \\ &= c_{1:k}\, L_{1:k}(\boldsymbol{\theta}). \end{aligned} \tag{32}$$

Equation (32) shows that realizations of $\boldsymbol{\Theta}$ with larger likelihood have a larger probability of belonging to the BUS event. Consequently, conditioning the augmented prior distribution of $(\boldsymbol{\Theta}, \Pi)$ on $\mathcal{O}_{1:k}$ weights parameter realizations according to their likelihood. This suggests that the resulting marginal conditional distribution of $\boldsymbol{\Theta}$ coincides with the Bayesian posterior.

The conditional density in the augmented space is

$$p(\boldsymbol{\theta}, \pi | \mathcal{O}_{1:k}) = \frac{p(\boldsymbol{\theta}, \pi) \, \mathbb{I}_{\mathcal{O}_{1:k}}(\boldsymbol{\theta}, \pi)}{\Pr(\mathcal{O}_{1:k})}, \tag{33}$$

where $\mathbb{I}_{\mathcal{O}_{1:k}}(\boldsymbol{\theta}, \pi)$ denotes the indicator function of the BUS event,

$$\mathbb{I}_{\mathcal{O}_{1:k}}(\boldsymbol{\theta}, \pi) = \begin{cases} 1, & (\boldsymbol{\theta}, \pi) \in \mathcal{O}_{1:k}, \\ 0, & \text{otherwise.} \end{cases}.$$

Since $\Pi$ is independent of $\boldsymbol{\Theta}$, the joint augmented prior density factorizes as $p(\boldsymbol{\theta}, \pi) = p(\boldsymbol{\theta}) \, p(\pi)$. Substituting into Eq. (33) gives

$$p(\boldsymbol{\theta}, \pi | \mathcal{O}_{1:k}) = \frac{p(\boldsymbol{\theta}) \, p(\pi) \, \mathbb{I}_{\mathcal{O}_{1:k}}(\boldsymbol{\theta}, \pi)}{\Pr(\mathcal{O}_{1:k})}. \tag{34}$$

The equivalence between the marginal BUS conditional distribution $p(\boldsymbol{\theta} | \mathcal{O}_{1:k})$ of $\boldsymbol{\Theta}$ and the Bayesian posterior $p(\boldsymbol{\theta} | \boldsymbol{z}_{1:k})$ is formalized in the following theorem.

---

**Theorem 1 (Equivalence between the BUS conditional distribution and Bayesian posterior).** The marginal distribution of $\boldsymbol{\Theta}$ conditional on the BUS event $\mathcal{O}_{1:k}$ is equal to the Bayesian posterior,

$$p(\boldsymbol{\theta} | \mathcal{O}_{1:k}) = p(\boldsymbol{\theta} | \boldsymbol{z}_{1:k}). \tag{35}$$

**Proof.** Applying Bayes' theorem gives

$$\Pr(\boldsymbol{\theta} | \mathcal{O}_{1:k}) = \frac{\Pr(\mathcal{O}_{1:k} | \boldsymbol{\Theta} = \boldsymbol{\theta}) \, p(\boldsymbol{\theta})}{\Pr(\mathcal{O}_{1:k})}.$$

Using Eq. (32), the probability of the BUS event is

$$\begin{aligned} \Pr(\mathcal{O}_{1:k}) &= \int \Pr(\mathcal{O}_{1:k} | \boldsymbol{\Theta} = \boldsymbol{\theta}) \, p(\boldsymbol{\theta}) \, d\boldsymbol{\theta} \\ &= c_{1:k} \int L_{1:k}(\boldsymbol{\theta}) \, p(\boldsymbol{\theta}) \, d\boldsymbol{\theta} \\ &= c_{1:k} \, c_{E,1:k}, \end{aligned} \tag{36}$$

where $c_{E,1:k}$ denotes the evidence introduced in Section 2.2. Hence,

$$p(\boldsymbol{\theta}|\mathcal{O}_{1:k}) = \frac{c_{1:k}\, L_{1:k}(\boldsymbol{\theta})\, p(\boldsymbol{\theta})}{c_{1:k}\, c_{E,1:k}}$$

$$= \frac{L_{1:k}(\boldsymbol{\theta})\, p(\boldsymbol{\theta})}{c_{E,1:k}}$$

$$= p(\boldsymbol{\theta}|\boldsymbol{z}_{1:k}),$$

which proves Eq. (35).

■

---

Theorem 1 shows that posterior samples can be obtained by sampling the augmented prior distribution conditional on $\mathcal{O}_{1:k}$ and retaining the corresponding parameter components. The Bayesian inference problem is therefore transformed into a conditional sampling problem in the augmented parameter space.

To express this problem in the form of a structural reliability problem, define the artificial limit-state function

$$g_O(\boldsymbol{\theta}, \pi, k) = \pi - c_{1:k}\, L_{1:k}(\boldsymbol{\theta}). \tag{37}$$

The BUS event in Eq. (31) can then equivalently be expressed as

$$\mathcal{O}_{1:k} = \{g_O(\boldsymbol{\Theta}, \Pi, k) \leq 0\}. \tag{38}$$

Bayesian updating is thereby reformulated as conditional sampling in an artificial failure-type domain of the augmented parameter space. According to Theorem 1, posterior samples are obtained by sampling the augmented prior distribution conditional on $\mathcal{O}_{1:k}$ and retaining the corresponding parameter components.

### 3.2 BUS with Subset Simulation

Direct Monte Carlo sampling conditional on $\mathcal{O}_{1:k}$ may become inefficient when the BUS event has a small probability under the augmented prior distribution. Subset Simulation [25] addresses this problem by reaching $\mathcal{O}_{1:k}$ through a sequence of more probable intermediate conditional events,

$$\mathcal{O}_{1:k}^{(1)} \supset \mathcal{O}_{1:k}^{(2)} \supset \cdots \supset \mathcal{O}_{1:k}{}^{(m)} = \mathcal{O}_{1:k}, \tag{39}$$

through decreasing thresholds

$$b_1 > b_2 > \cdots > b_{m-1} > b_m = 0 \tag{40}$$

such that

$$\mathcal{O}_{1:k}^{(j)} = \{g_O(\boldsymbol{\Theta}, \Pi, k) \leq b_j\}, \quad j = 1, \dots, m. \tag{41}$$

Since the events are nested, the probability of the BUS event can be decomposed as

$$\Pr(\mathcal{O}_{1:k}) = \Pr(\mathcal{O}_{1:k}^{(1)}) \prod_{j=2}^{m} \Pr(\mathcal{O}_{1:k}^{(j)} | \mathcal{O}_{1:k}^{(j-1)}) \tag{42}$$

The intermediate thresholds $b_j$ are selected adaptively such that approximately a prescribed fraction $p_0$ of the samples at one level belongs to the next event. These samples are retained as seeds, and MCMC is used to generate additional samples conditional on the new event. Thus, at Subset Simulation level $j$, the samples target

$$p(\boldsymbol{\theta}, u | \mathcal{O}_{1:k}^{(j)}). \tag{43}$$

The procedure is repeated until the threshold reaches $b_m = 0$, for which $p(\boldsymbol{\theta}, u | \mathcal{O}_{1:k})$, and, according to Theorem 1, their parameter components follow the Bayesian posterior, $p(\boldsymbol{\theta} | \boldsymbol{z}_{1:k})$. The procedure is summarized in Algorithm 1.

---

**Algorithm 1: BUS with Subset Simulation [24]**

**Input:** prior distribution $p(\boldsymbol{\theta})$, monitoring stage $k$, cumulative likelihood $L_{1:k}(\boldsymbol{\theta})$, likelihood multiplier $c_{1:k}$ satisfying $c_{1:k}\, L_{1:k}(\boldsymbol{\theta}) \leq 1$, sample size $N$, and target conditional probability $p_0$.

Define

1. **Define the BUS limit-state function:** Define $g_O(\boldsymbol{\theta}, u, k) = \pi - c_{1:k}\, L_{1:k}(\boldsymbol{\theta})$.
2. **Generate initial samples:** Generate $N$ independent samples $\boldsymbol{\theta}^{(i)} \sim p(\boldsymbol{\theta})$, $\pi^{(i)} \sim \mathcal{U}(0,1)$, $i = 1, \dots, N$, and evaluate $g_O(\boldsymbol{\theta}^{(i)}, \pi^{(i)}, k)$.
3. **Generate intermediate conditional samples:** At level $j$, determine a threshold $b_j > 0$ such that approximately a fraction $p_0$ of the samples satisfies $g_O(\boldsymbol{\theta}, \pi, k) \leq b_j$. These samples belong to $\mathcal{O}_{1:k}^{(j)} = \{g_O(\boldsymbol{\Theta}, \Pi, k) \leq b_j\}$. Retain them as seeds and use MCMC to generate $N$ samples from $p(\boldsymbol{\theta}, \pi | \mathcal{O}_{1:k}^{(j)})$. Repeat this step until the next intermediate threshold would be non-positive.
4. **Generate sample conditional on the BUS event:** Set the final threshold to $b_m = 0$, such that $\mathcal{O}_{1:k}^{(m)} = \mathcal{O}_{1:k}$. Use the samples satisfying the BUS event as seeds and apply MCMC to generate $N$ samples from $p(\boldsymbol{\theta}, \pi | \mathcal{O}_{1:k})$.
5. **Obtain posterior samples:** Retain the final augmented samples $(\boldsymbol{\theta}^{(i)}, \pi^{(i)}) \sim p(\boldsymbol{\theta}, \pi | \mathcal{O}_{1:k})$. According to Theorem 1, their parameter components satisfy $\boldsymbol{\theta}^{(i)} \sim p(\boldsymbol{\theta} | \boldsymbol{z}_{1:k})$.

6. **Estimate the BUS-event probability and evidence.** Estimate the probability of the BUS event from the Subset Simulation levels as $\widehat{\Pr}(\mathcal{O}_{1:k}) = \widehat{\Pr}(\mathcal{O}_{1:k}^{(1)}) \prod_{j=2}^{m} \widehat{\Pr}(\mathcal{O}_{1:k}^{(j)} | \mathcal{O}_{1:k}^{(j-1)})$, with $\mathcal{O}_{1:k}^{(m)} = \mathcal{O}_{1:k}$. For intermediate levels selected with target conditional probability $p_0$, this becomes $\widehat{\Pr}(\mathcal{O}_{1:k}) = p_0^{m-1}\ \widehat{\Pr}(\mathcal{O}_{1:k} | \mathcal{O}_{1:k}^{(m-1)})$, where the final factor is estimated from the fraction of samples at the last intermediate level that belong to $\mathcal{O}_{1:k}$. Using the relation provided by Eq. (36), estimate the evidence as $\hat{c}_{E,1:k} = \widehat{\Pr}(\mathcal{O}_{1:k}) / c_{1:k}$.

---

BUS with Subset Simulation therefore replaces direct sampling of a potentially rare BUS event by a sequence of more probable conditional sampling problems. Importantly, the result is not only a set of posterior parameter samples, but a sample population from the augmented conditional distribution $p(\boldsymbol{\theta}, \pi | \mathcal{O}_{1:k})$. This augmented conditional sample population can be retained for subsequent Bayesian updating as new monitoring data become available.

### 3.3 Sequential BUS

Section 2.2 formulated sequential Bayesian inference as the recursive update of the posterior distribution when new monitoring data become available. The augmented conditional sample population obtained with BUS provides a natural basis for implementing this update without restarting the inference from the prior distribution.

Consider the transition from monitoring stage $k-1$ to stage $k$. At monitoring stage $k-1$, the available augmented samples obtained with BUS follow $p(\boldsymbol{\theta}, \pi | \mathcal{O}_{1:k-1})$, and, according to Theorem 1, their parameter components follow $p(\boldsymbol{\theta} | \boldsymbol{z}_{1:k-1})$. Importantly, the posterior from the preceding monitoring stage does not have to be introduced as a new explicit sampling distribution. Its augmented representation is given by the augmented prior distribution restricted to the preceding BUS event,

$$p(\boldsymbol{\theta}, \pi | \mathcal{O}_{1:k-1}) = \frac{p(\boldsymbol{\theta})\, p(\pi)\, \mathbb{I}_{\mathcal{O}_{1:k-1}}(\boldsymbol{\theta}, \pi)}{\Pr(\mathcal{O}_{1:k-1})}.$$

Sequential BUS exploits this representation to incorporate new monitoring data $\mathbf{z}_k$ by further conditioning the existing augmented sample population, rather than restarting the BUS analysis from the augmented prior distribution.

At monitoring stage $k-1$, the BUS construction of Section 3.1 defines the limit-state function $g_{\mathcal{O}}(\boldsymbol{\theta}, \pi, k-1)$ and corresponding BUS event $\mathcal{O}_{1:k-1}$ based on the cumulative likelihood $L_{1:k-1}(\boldsymbol{\theta})$. When the new monitoring data $\mathbf{z}_k$ become available, the cumulative likelihood is updated to $L_{1:k}(\boldsymbol{\theta})$, defining the new limit-state function $g_{\mathcal{O}}(\boldsymbol{\theta}, \pi, k)$ and BUS event $\mathcal{O}_{1:k}$.

The same auxiliary variable $\Pi$ and the same augmented prior distribution are retained across monitoring stages. To allow the augmented sample population from stage $k-1$ to serve directly as the starting population for the next update, the new BUS event is constructed such that $\mathcal{O}_{1:k} \subseteq \mathcal{O}_{1:k-1}$.

To ensure this nestedness, an incremental likelihood multiplier $c_k > 0$ is introduced such that $c_k\, L_{k|1:k-1}(\boldsymbol{\theta}) \leq 1$ for all $\boldsymbol{\theta}$, where $L_{k|1:k-1}(\boldsymbol{\theta})$ is the conditional likelihood defined in Eq. (15). The cumulative likelihood multiplier is defined recursively as $c_{1:k} = c_{1:k-1}\, c_k$. The resulting nested-event property is stated in the following proposition.

---

**Proposition 1 (Nestedness of successive BUS events).**

If

$$c_k\, L_{k|1:k-1}(\boldsymbol{\theta}) \leq 1, \quad \forall \boldsymbol{\theta}$$

and

$$c_{1:k} = c_{1:k-1}\, c_k$$

then

$$\mathcal{O}_{1:k} \subseteq \mathcal{O}_{1:k-1}.$$

**Proof.** Using the recursive likelihood relation of Eq. (17),

$$L_{1:k}(\boldsymbol{\theta}) = L_{1:k-1}(\boldsymbol{\theta})\, L_{k|1:k-1}(\boldsymbol{\theta}),$$

together with the recursive definition of $c_{1:k}$, gives

$$\begin{aligned} c_{1:k}\, L_{1:k}(\boldsymbol{\theta}) &= c_{1:k-1}\, c_k\, L_{1:k-1}(\boldsymbol{\theta})\, L_{k|1:k-1}(\boldsymbol{\theta}) \\ &= c_{1:k-1}\, L_{1:k-1}(\boldsymbol{\theta})\left[c_k\, L_{k|1:k-1}(\boldsymbol{\theta})\right] \\ &\leq c_{1:k-1}\, L_{1:k-1}(\boldsymbol{\theta}). \end{aligned}$$

Hence,

$$g_O(\boldsymbol{\theta}, \pi, k) \geq g_O(\boldsymbol{\theta}, \pi, k-1).$$

Therefore,

$$g_O(\boldsymbol{\theta}, \pi, k) \leq 0 \quad \Rightarrow \quad g_O(\boldsymbol{\theta}, \pi, k-1) \leq 0,$$

which proves

$$\mathcal{O}_{1:k} \subseteq \mathcal{O}_{1:k-1}.$$

■

---

Proposition 1 provides the key computational property of Sequential BUS. Since the available augmented samples already follow $p(\boldsymbol{\theta}, \pi|\mathcal{O}_{1:k-1})$, and $\mathcal{O}_{1:k} \subseteq \mathcal{O}_{1:k-1}$, the new posterior can be obtained by further conditioning this population on $\mathcal{O}_{1:k}$. The augmented prior distribution therefore does not need to be sampled again.

The corresponding posterior equivalence follows immediately.

---

**Corollary 1 (Sequential posterior equivalence).**

Further conditioning the augmented distribution at monitoring stage $k-1$ on $\mathcal{O}_{1:k}$ yields

$$p(\boldsymbol{\theta}|\mathcal{O}_{1:k-1}, \mathcal{O}_{1:k}) = p(\boldsymbol{\theta}|\boldsymbol{z}_{1:k}).$$

**Proof.** Since $\mathcal{O}_{1:k} \subseteq \mathcal{O}_{1:k-1}$,

$$p(\boldsymbol{\theta}|\mathcal{O}_{1:k-1}, \mathcal{O}_{1:k}) = p(\boldsymbol{\theta}|\mathcal{O}_{1:k}).$$

According to Theorem 1,

$$p(\boldsymbol{\theta}|\mathcal{O}_{1:k}) = p(\boldsymbol{\theta}|\boldsymbol{z}_{1:k}),$$

and therefore

$$p(\boldsymbol{\theta}|\mathcal{O}_{1:k-1}, \mathcal{O}_{1:k}) = p(\boldsymbol{\theta}|\boldsymbol{z}_{1:k}).$$

■

---

The nested construction also relates the conditional probability of the new BUS event to the evidence of the sequential Bayesian update. Since $\mathcal{O}_{1:k} \subseteq \mathcal{O}_{1:k-1}$, $\Pr(\mathcal{O}_{1:k}|\mathcal{O}_{1:k-1}) = \Pr(\mathcal{O}_{1:k}) / \Pr(\mathcal{O}_{1:k-1})$. Using $\Pr(\mathcal{O}_{1:k}) = c_{1:k}\, c_{E,1:k}$, together with $c_{E,1:k} = c_{E,1:k-1}\, c_{E,k|1:k-1}$ as defined in Eq. (20), gives

$$\Pr(\mathcal{O}_{1:k}|\mathcal{O}_{1:k-1}) = c_k\, c_{E,k|1:k-1}. \tag{44}$$

Sequential BUS thus represents successive Bayesian updates through a nested sequence of BUS events, $\mathcal{O}_{1:1} \supseteq \mathcal{O}_{1:2} \supseteq \cdots \supseteq \mathcal{O}_{1:k}$. Each event is defined using the cumulative likelihood and therefore represents all monitoring information available up to the respective stage. The nested construction makes it possible to carry the augmented conditional sample population from one monitoring stage to the next. Section 3.4 describes how this sequential conditioning is implemented with Subset Simulation, starting from the augmented sample population conditional on $\mathcal{O}_{1:k-1}$ and progressing toward $\mathcal{O}_{1:k}$.

### 3.4 Sequential BUS with subset simulation

The nestedness established in Proposition 1 provides the basis for implementing Sequential BUS with Subset Simulation. After monitoring stage $k-1$, augmented samples are available from the distribution conditional on $\mathcal{O}_{1:k-1}$.

At monitoring stage $k$, the cumulative likelihood and likelihood multiplier are given by $L_{1:k}(\boldsymbol{\theta}) = L_{1:k-1}(\boldsymbol{\theta}) L_{k|1:k-1}(\boldsymbol{\theta})$ and $c_{1:k} = c_{1:k-1}\, c_k$, respectively, where $c_k$ is selected such that $c_k\, L_{k|1:k-1}(\theta) \leq 1$ for all $\boldsymbol{\theta}$. The corresponding BUS limit-state function is $g_O(\boldsymbol{\theta}, \pi, k) = \pi - c_{1:k}\, L_{1:k}(\boldsymbol{\theta})$, and the new BUS event is $\mathcal{O}_{1:k} = \{g_O(\boldsymbol{\Theta}, \Pi, k) \leq 0\}$. According to Proposition 1, $\mathcal{O}_{1:k} \subseteq \mathcal{O}_{1:k-1}$.

The transition from $\mathcal{O}_{1:k-1}$ to $\mathcal{O}_{1:k}$ is performed by Subset Simulation while retaining the preceding BUS event as a fixed conditioning domain, consistent with the conditional Subset Simulation formulations in Straub et al. [26] and Jerez et al. [28]. The initial event for the Subset Simulation procedure is therefore defined as

$$\mathcal{O}_{1:k}^{(0)} = \mathcal{O}_{1:k-1}. \tag{45}$$

At subsequent levels $j = 1, \dots, m$, the intermediate events are defined as

$$\mathcal{O}_{1:k}^{(j)} = \mathcal{O}_{1:k-1} \cap \{g_O(\boldsymbol{\Theta}, \Pi, k) \leq b_j\}, \quad j = 1, \dots, m. \tag{46}$$

where the intermediate thresholds satisfy $b_1 > b_2 > \cdots > b_m = 0$. Consequently,

$$\mathcal{O}_{1:k-1} = \mathcal{O}_{1:k}^{(0)} \supset \mathcal{O}_{1:k}^{(1)} \supset \cdots \supset {\mathcal{O}_{1:k}}^{(m)} = \mathcal{O}_{1:k}. \tag{47}$$

At each intermediate level $j$, the threshold $b_j$ is selected adaptively such that approximately a prescribed fraction $p_0$ of the samples conditional $\mathcal{O}_{1:k}^{(j-1)}$ satisfies $g_O(\boldsymbol{\theta}, \pi, k) \leq b_j$. Since $\mathcal{O}_{1:k}^{(j-1)} \subseteq \mathcal{O}_{1:k-1}$, these samples constitute seeds in $\mathcal{O}_{1:k}^{(j)}$. Conditional MCMC is subsequently used to generate samples from the distribution $p(\boldsymbol{\theta}, \pi | \mathcal{O}_{1:k}^{(j)}) = p(\boldsymbol{\theta}, \pi | \mathcal{O}_{1:k-1} \cap \{g_O(\boldsymbol{\Theta}, \Pi, k) \leq b_j\})$. Accordingly, the event-membership test within the conditional sampler requires a proposed state $(\boldsymbol{\theta}, \pi)$ must satisfy both $g_O(\boldsymbol{\theta}, \pi, k) \leq b_j$ and $g_O(\boldsymbol{\theta}, \pi, k-1) \leq 0$. The preceding BUS event therefore remains a fixed constraint throughout the transition from $\mathcal{O}_{1:k-1}$ to $\mathcal{O}_{1:k}$. Apart from this additional constraint, the adaptive selection of the intermediate thresholds and the generation of conditional samples follow the Subset Simulation procedure described in Section 3.2.

The procedure continues until the threshold reaches zero. At the final level,

$$
\begin{aligned}
\mathcal{O}_{1:k}^{(m)} &= \mathcal{O}_{1:k-1} \cap \{g_{\mathcal{O}}(\boldsymbol{\Theta}, \Pi, k) \leq 0\} \\
&= \mathcal{O}_{1:k-1} \cap \mathcal{O}_{1:k}.
\end{aligned}
\tag{48}
$$

Since Proposition 1 establishes that $\mathcal{O}_{1:k} \subseteq \mathcal{O}_{1:k-1}$, it follows that $\mathcal{O}_{1:k}^{(m)} = \mathcal{O}_{1:k}$.

The resulting augmented samples therefore follow $p(\boldsymbol{\theta}, \pi \mid \mathcal{O}_{1:k})$, and, according to Theorem 1, their $\boldsymbol{\theta}$-components follow the Bayesian posterior conditional on all monitoring information available up to stage $k$. The complete augmented sample population, including the auxiliary variable $\pi$, is retained for the subsequent monitoring update.

The conditional probability of the new BUS event can be expressed as the product of the conditional probabilities associated with the intermediate events,

$$
\Pr(\mathcal{O}_{1:k} | \mathcal{O}_{1:k-1}) = \prod_{j=1}^{m} \Pr(\mathcal{O}_{1:k}^{(j)} | \mathcal{O}_{1:k}^{(j-1)}) \tag{49}
$$

For the intermediate levels, the thresholds are selected such that the corresponding conditional probabilities are approximately $p_0$. At the final level, the conditional probability is estimated from the fraction of samples at the preceding level satisfying $g_{\mathcal{O}}(\boldsymbol{\theta}, \pi, k) \leq 0$. The corresponding conditional evidence of the sequential Bayesian update is obtained from the estimated conditional BUS-event probability according to Eq. (44).

---

**Algorithm 2: Sequential BUS with Subset Simulation**

**Input:** monitoring stage $k$, augmented samples conditional on $\mathcal{O}_{1:k-1}$, cumulative likelihood $L_{1:k-1}(\boldsymbol{\theta})$, cumulative likelihood multiplier $c_{1:k-1}$, incremental likelihood $L_{k|1:k-1}(\boldsymbol{\theta})$, incremental likelihood multiplier $c_k$ satisfying $c_k\, L_{k|1:k-1}(\boldsymbol{\theta}) \leq 1$ for all $\boldsymbol{\theta}$, sample size $N$, and target conditional probability $p_0$.

Define

1. **Define the preceding and updated BUS limit-state functions:** Define the preceding BUS limit-state function as $g_{\mathcal{O}}(\boldsymbol{\theta}, \pi, k-1) = \pi - c_{1:k-1}\, L_{1:k-1}(\boldsymbol{\theta})$, such that $\mathcal{O}_{1:k-1} = \{g_{\mathcal{O}}(\boldsymbol{\Theta}, \Pi, k-1) \leq 0\}$. Update the cumulative likelihood and likelihood multiplier as $L_{1:k}(\boldsymbol{\theta}) = L_{1:k-1}(\boldsymbol{\theta})\, L_{k|1:k-1}(\boldsymbol{\theta})$ and $c_{1:k} = c_{1:k-1}\, c_k$ and define the new BUS limit-state function $g_{\mathcal{O}}(\boldsymbol{\theta}, \pi, k) = \pi - c_{1:k}\, L_{1:k}(\boldsymbol{\theta})$. Evaluate $g_{\mathcal{O}}(\boldsymbol{\theta}, \pi, k)$ for the augmented samples conditional on $\mathcal{O}_{1:k-1}$.
2. **Generate intermediate conditional samples:** For notational convenience, let $\mathcal{O}_{1:k}^{(0)} = \mathcal{O}_{1:k-1}$. At level $j$, determine $b_j > 0$ such that approximately a fraction $p_0$ of the current samples satisfies $g_{\mathcal{O}}(\boldsymbol{\theta}, \pi, k) \leq b_j$. Define $\mathcal{O}_{1:k}^{(j)} = \mathcal{O}_{1:k-1} \cap \{g_{\mathcal{O}}(\boldsymbol{\Theta}, \Pi, k) \leq b_j\}$. Retain the samples belonging to $\mathcal{O}_{1:k}^{(j)}$ as seeds and use conditional MCMC to generate $N$ samples from $p(\boldsymbol{\theta}, \pi | \mathcal{O}_{1:k}^{(j)})$. During conditional sampling, proposed states must satisfy both the current intermediate threshold and the preceding BUS-event constraint, $g_O(\boldsymbol{\theta}, \pi, k) \leq b_j$, $g_O(\boldsymbol{\theta}, \pi, k-1) \leq 0$. Repeat until the next threshold would be non-positive.

3. **Generate samples conditional on the updated BUS event:** Set the final threshold to $b_m = 0$. The final intermediate event is therefore $\mathcal{O}_{1:k}^{(m)} = \mathcal{O}_{1:k-1} \cap \{g_O(\boldsymbol{\Theta}, \Pi, k) \leq 0\} = \mathcal{O}_{1:k-1} \cap \mathcal{O}_{1:k} = \mathcal{O}_{1:k}$, where the final equality follows from Proposition 1. Retain the samples satisfying $\mathcal{O}_{1:k}$as seeds and use conditional MCMC to generate $N$samples from $p(\boldsymbol{\theta}, \pi | \mathcal{O}_{1:k})$.
4. **Obtain the updated samples:** Retain the resulting augmented conditional on $\mathcal{O}_{1:k}$. According to Theorem 1, their parameter components follow the Bayesian posterior associated with the monitoring information up to stage $k$. The augmented sample population is retained for the subsequent monitoring stage.
5. **Estimate the conditional BUS-event probability and evidence.** Estimate $\widehat{\Pr}(\mathcal{O}_{1:k} | \mathcal{O}_{1:k-1}) = \prod_{j=1}^{m} \widehat{\Pr}(\mathcal{O}_{1:k}^{(j)} | \mathcal{O}_{1:k}^{(j-1)})$. For intermediate levels selected with target conditional probability $p_0$, the corresponding factors are approximately $p_0$, while the final factor is estimated from the fraction of samples at the last intermediate level that satisfy the updated BUS event. Using Eq. (44), estimate the corresponding conditional evidence.

---

Sequential BUS with Subset Simulation therefore extends the conditional-sampling sequence across monitoring stages. The adaptive threshold-selection principle of Subset Simulation remains unchanged, while conditional MCMC at each intermediate level is additionally restricted to the preceding BUS event $\mathcal{O}_{1:k-1}$. Consequently, the augmented conditional sample population obtained at the preceding monitoring stage replaces the augmented prior population as the starting point for the transition toward the new nested BUS event.

# 4 Illustrative example

The proposed framework is illustrated using the jacket-type tubular steel frame structure shown in Figure 1, considering fatigue deterioration at its welded brace-to-chord connections. The structure originates from an experimental research program conducted at the University of California, Berkeley in the 1980s to investigate the cyclic and inelastic behavior of braced frame structures [32]. Since then, it has become a widely used benchmark in structural engineering, with applications including the verification of nonlinear static pushover analysis methods (e.g., [33]), vibration-based damage detection (e.g., [34]), system reliability analysis and updating of deteriorating structural systems (e.g., [13]), and risk-based inspection and maintenance planning (e.g., [35]). It therefore provides a well-established structural example for illustrating the proposed framework.

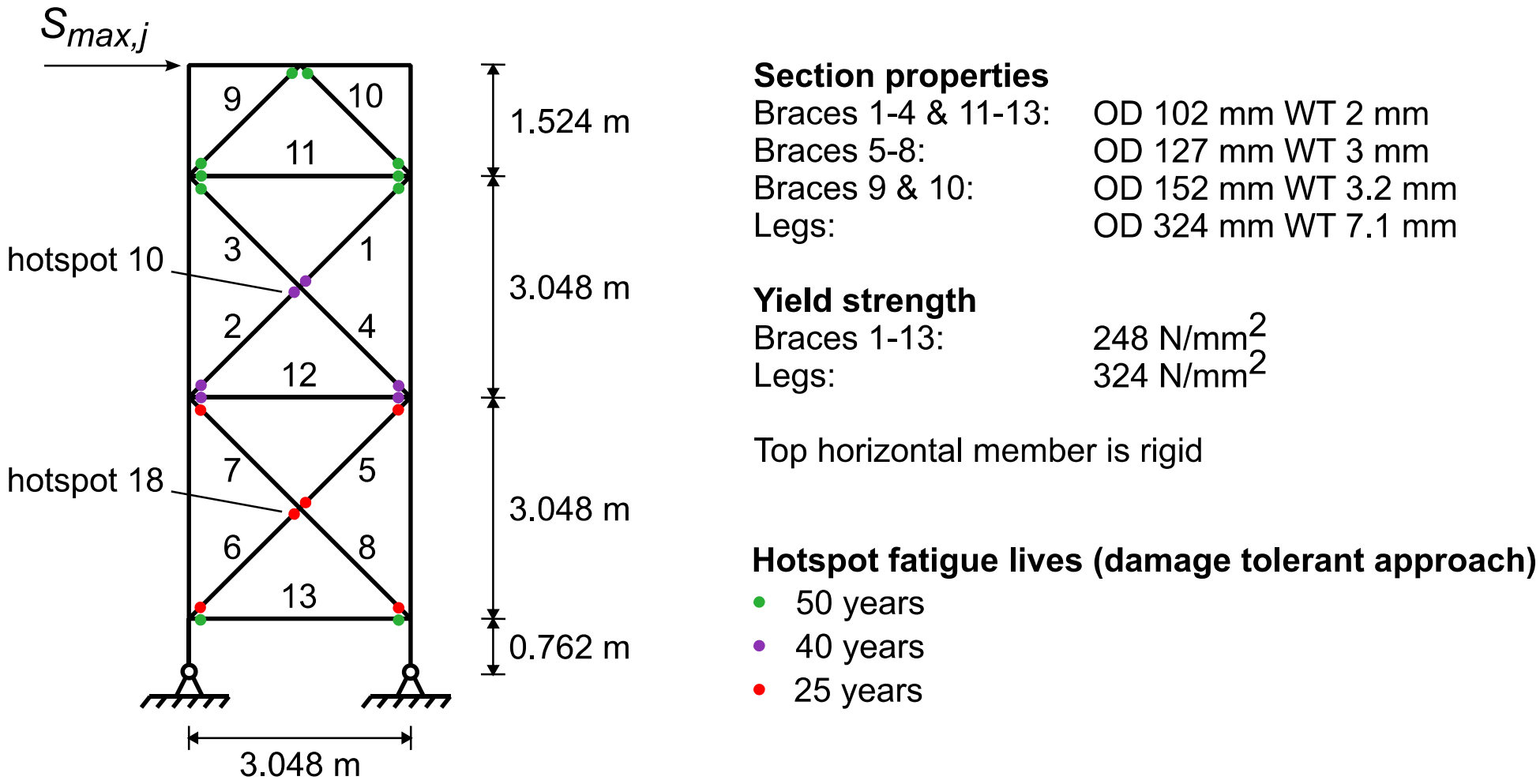

**Figure 1:** Geometry, braces and fatigue-hotspots, structural properties, hotspot fatigue-life classes, and structural loading of the Zayas frame considered in the illustrative example.

The example considers a 25-year service life during which global dynamic monitoring is performed annually. The monitoring data are used sequentially to update probabilistic fatigue deterioration models and subsequently to perform system-level damage diagnosis, predict hotspot deterioration and fatigue failure, and update the structural reliability. In addition to demonstrating these monitoring-informed engineering predictions, the example illustrates how the deterioration processes relevant to structural performance, the required resolution of the deterioration model, and the selection of the monitoring strategy are interrelated. These considerations are first discussed before the deterioration and monitoring models are formulated. The probabilistic model specifications, numerical parameter values, and computational settings used throughout the illustrative example are summarized in Appendix A.

## 4.1 Structural system and monitoring problem definition

Fatigue is considered the relevant deterioration mechanism for the welded brace-to-chord connections of the structure. During most of the fatigue crack-growth process, a surface-breaking crack is assumed to have negligible influence on both the stiffness and the load-carrying capacity of the connection. This is consistent with conventional fatigue design, in which fatigue life is commonly associated with crack growth up to the section thickness, and with experimental observations showing that local surface cracks do not produce a measurable change in the global structural behavior [6]. Once the fatigue crack at a hotspot has grown through the section thickness, however, the corresponding welded connection is assumed to unzip rapidly and sever, such that the associated brace no longer contributes to the structural stiffness or resistance. A through-thickness crack is therefore taken as the transition from continuous local fatigue deterioration to failure of the corresponding connection.

At the structural-system level, it is therefore the failure of a connection, rather than the preceding growth of a surface crack, that produces a relevant change in structural behavior and performance.

The loss of the associated brace reduces the global stiffness and can reduce the ultimate structural capacity. Owing to the redundancy of the frame, however, failure of an individual connection does not necessarily lead to immediate system collapse. This makes the structure well suited for the development of monitoring-informed lifetime-management strategies based on global dynamic monitoring, since connection failures can be identified through the resulting changes in the global dynamic properties. Such a degraded state may be unacceptable for long-term operation, while still providing a time window for implementing repair or other maintenance actions.

These considerations determine the required level of detail of the deterioration model. The local fatigue process must be represented with sufficient resolution to predict when a hotspot crack reaches the section thickness and hence triggers the transition to the failed connection state. At the same time, no detailed representation of the influence of subcritical crack growth on the global structural model is required, since such cracks are assumed not to alter the global stiffness or resistance before through-thickness cracking. The deterioration model introduced in Section 4.2 therefore combines continuous crack-growth models at the hotspot level with a discrete representation of connection and system damage states.

The monitoring setup is selected consistently with these system-level consequences of deterioration. Since subcritical surface cracks do not produce an appreciable change in global structural behavior, they are not expected to be observable by global dynamic monitoring. Connection failures, in contrast, remove the stiffness contribution of individual braces and thereby affect the global dynamic properties of the structure. Global dynamic monitoring is therefore adopted to provide information on the discrete connection and system damage states rather than to observe the local crack-growth process directly. The corresponding monitoring setup and probabilistic monitoring model are introduced in Section 4.3.

## 4.2 Fatigue deterioration and system deterioration model

Fatigue deterioration is modeled at the critical hotspots of the 22 welded connections shown as colored dots in Figure 1. For each connection $j$, the surface-crack depth $A_j(t) = a_j(t; \boldsymbol{\Theta}_D)$ is described by a one-dimensional Paris-law model as

$$a_j(t; \boldsymbol{\Theta}_D) = \left[\left(1 - \frac{m_j}{2}\right) C_j \, Y_j^{m_j} \, B_{SIF,j}^{m_j} \, B_{S,j}^{m_j} \, \Delta S_{e,j}^{m_j} \, \pi^{m_j/2} \, \nu \, t + A_{0,j}^{(1-m_j/2)}\right]^{(1-m_j/2)^{-1}} \tag{50}$$

with

$$\Delta S_{e,j} = k_j \, \Gamma\left(1 + \frac{m_j}{\lambda_j}\right)^{1/m_j}. \tag{51}$$

Here, $A_{0,j}$ is the initial crack depth, $C_j$ and $m_j$ are the Paris-law parameters, $Y_j$ is the geometry factor entering the stress-intensity-factor model, $B_{S,j}$ and $B_{\mathrm{SIF},j}$ account for model uncertainty in the fatigue-load and stress-intensity-factor models, respectively, and $\Delta S_{e,j}$ is the equivalent stress range associated with the Weibull stress-range model. The uncertain deterioration parameters are collected in

$$\boldsymbol{\Theta}_D = [B_{SIF,1}, B_{S,1}, C_1, A_{0,1}, \ldots, B_{SIF,22}, B_{S,22}, C_{22}, A_{0,22}]^T \tag{52}$$

such that $\boldsymbol{\Theta}_D$ comprises 88 random variables in total.

The prior fatigue model is constructed from the available design information following [36]. The marginal prior distributions of the fatigue-model parameters are specified at the hotspot level. In particular, the hotspot-specific stress-range scale parameters $k_j$ are determined from the design fatigue lives shown in Figure 1, the design S–N curve, and the associated partial safety factors. Selected fracture-mechanics parameters are subsequently calibrated to the corresponding probabilistic S–N model such that the two representations provide consistent fatigue reliabilities. Dependence between hotspots is introduced through estimated common inter-hotspot correlation coefficients for the corresponding fatigue-model parameters.

A connection is assumed to retain its structural contribution until the crack reaches the critical depth $a_{c,j}$. Once this condition is reached, the joint is assumed to unzip rapidly and sever, such that the associated brace no longer contributes to structural stiffness or resistance. The corresponding discrete joint state is therefore defined as

$$J_j(t) = \begin{cases} 0, & A_j(t) < a_{c,j}, \\ 1, & \text{otherwise}, \end{cases} \tag{53}$$

where $J_j(t) = 0$ and $J_j(t) = 1$ denote an intact and failed joint, respectively. The state of brace $i$ is subsequently obtained as

$$D_i(t) = \max_{j \in \mathcal{J}_i} J_j(t), \tag{54}$$

where $\mathcal{J}_i$ denotes the set of fatigue-critical joints associated with brace $i$. The system deterioration state is therefore

$$\boldsymbol{D}(t) = [D_1(t), \ldots, D_{n_b}(t)]^T, \tag{55}$$

where $n_b = 13$ is the number of braces in the structural system. In accordance with the general formulation in Section 2.1, the complete deterioration model can therefore be written as $\boldsymbol{D}(t) = \boldsymbol{h}_D(t; \boldsymbol{\Theta}_D)$, where $\boldsymbol{h}_D$ comprises the crack-growth model, the joint-failure criterion, and the mapping from joint failures to brace failures. Thus,

$$\boldsymbol{\Theta}_D \rightarrow \{A_j(t)\} \rightarrow \{J_j(t)\} \rightarrow \boldsymbol{D}(t). \tag{56}$$

### 4.3 Monitoring setup and monitoring model

The monitoring setup consists of horizontal acceleration sensors located as shown in Figure 2. At each monitoring stage $k$, the measured vibration responses are processed using operational modal analysis (OMA) to identify $n_m$ modal eigenvalues $\hat{\lambda}_{k,r}$ and corresponding mode shapes $\widehat{\mathbf{\Phi}}_{k,r}$, $r = 1, \dots, n_m$. The observation vector introduced in Section 2 is therefore

$$\boldsymbol{z}_k = [\hat{\lambda}_{k,1}, \dots, \hat{\lambda}_{k,n_m}, \widehat{\mathbf{\Phi}}_{k,1}^T, \dots, \widehat{\mathbf{\Phi}}_{k,n_m}^T]^T. \tag{57}$$

The modal properties provide indirect information on fatigue deterioration because brace failures alter the structural stiffness and, consequently, the dynamic characteristics of the structure.

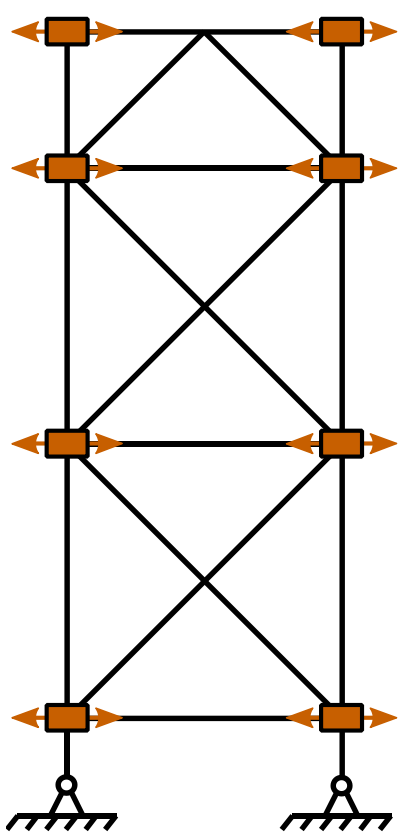

**Figure 2:** Monitoring setup with horizontal acceleration sensors used for the annual vibration measurements.

A linear-elastic FE model implemented in OpenSees [37, 38] provides the structural behavior model $\mathbf{h}_B$ introduced in Section 2.1. For given model parameters and deterioration state, the FE model provides deterministic predictions of the modal properties. In the present example, the FE-model parameters are fixed at values that are assumed to have been selected based on a preceding system-identification step. Accordingly, the structural-behavior parameters $\mathbf{\Theta}_B$ of the general formulation in Section 2 do not appear as arguments in the following expressions. For a realization $\boldsymbol{\theta}_D$, the deterioration model of Section 4.2 determines the brace-level deterioration state at monitoring time $t_k$. Failed braces are removed from the FE model, and eigenvalue analysis yields the corresponding model-predicted eigenvalues $\lambda_{k,r}(\boldsymbol{\theta}_D)$ and mode shapes $\mathbf{\Phi}_{k,r}(\boldsymbol{\theta}_D)$. After pairing the numerical and identified modes as described below, the structural behavior vector of Section 2 becomes

$$\boldsymbol{y}_k(\boldsymbol{\theta}_D) = \boldsymbol{h}_B(\boldsymbol{h}_D(t_k; \boldsymbol{\Theta}_D)) = \left[\lambda_{k,1}(\boldsymbol{\theta}_D), \dots, \lambda_{k,n_m}(\boldsymbol{\theta}_D), \mathbf{\Phi}_{k,1}^T(\boldsymbol{\theta}_D), \dots, \mathbf{\Phi}_{k,n_m}^T(\boldsymbol{\theta}_D)\right]^T. \tag{58}$$

where $\mathbf{\Phi}_{k,r}(\boldsymbol{\theta}_D)$ contains only the mode-shape coordinates associated with the instrumented degrees of freedom.

Because the ordering of identified and model-predicted modes may differ, mode matching is performed for each realization $\boldsymbol{\theta}_D$ before evaluating the likelihood. The similarity between identified mode $r$ and candidate numerical mode $q$ is quantified by the modal assurance criterion (MAC)

$$\mathrm{MAC}_{k,rq}(\boldsymbol{\theta}_D) = \frac{(\widehat{\boldsymbol{\Phi}}_{k,r}^T\,\boldsymbol{\Phi}_{k,q}(\boldsymbol{\theta}_D))^2}{\|\widehat{\boldsymbol{\Phi}}_{k,r}\|^2\|\boldsymbol{\Phi}_{k,q}(\boldsymbol{\theta}_D)\|^2} \tag{59}$$

Following Simoen et al. [29], the MAC and the corresponding eigenvalue ratio are combined in the matching criterion

$$\delta_{k,rq} = 1 - \mathrm{MAC}_{k,rq}(\boldsymbol{\theta}_D) + |1 - \hat{\lambda}_{k,r}/\lambda_{k,q}(\boldsymbol{\theta}_D)|. \tag{60}$$

The modes are matched iteratively according to the smallest values of $\delta_{k,rq}$ until each identified mode has been assigned a numerical counterpart. After matching, the model-predicted quantities are indexed by the corresponding identified mode $r$.

Following Vanik et al. [18], Gaussian prediction errors are adopted for the matched modal quantities. For eigenvalue $r$ at monitoring stage $k$, the prediction error is

$$\varepsilon_{\lambda,k,r}(\boldsymbol{\theta}_D) = \hat{\lambda}_{k,r} - \lambda_{k,r}(\boldsymbol{\theta}_D), \tag{61}$$

with standard deviation

$$\sigma_{\lambda,k,r} = c_{\lambda,r}\,\hat{\lambda}_{k,r}. \tag{62}$$

For the mode shapes, let $\hat{\phi}_{k,r,s}$ and $\phi_{k,r,s}(\boldsymbol{\theta}_D)$ denote the $s$th coordinates of the identified and model-predicted mode shapes $\widehat{\boldsymbol{\Phi}}_{k,r}$ and $\boldsymbol{\Phi}_{k,r}(\boldsymbol{\theta}_D)$, respectively. The prediction error of each mode-shape coordinate is then

$$\varepsilon_{\Phi,k,r,s}(\boldsymbol{\theta}_D) = \hat{\phi}_{k,r,s} - \gamma_{k,r}(\boldsymbol{\theta}_D)\,\phi_{k,r,s}(\boldsymbol{\theta}_D), \quad s = 1, \dots, n_\Phi, \tag{63}$$

with standard deviation

$$\sigma_{\Phi,k,r} = c_{\Phi,r}\,\|\widehat{\boldsymbol{\Phi}}_{k,r}\|, \tag{64}$$

where $n_\Phi$ denotes the number of considered mode-shape coordinates and

$$\gamma_{k,r}(\boldsymbol{\theta}_D) = \frac{\widehat{\boldsymbol{\Phi}}_{k,r}^T\,\boldsymbol{\Phi}_{k,q}(\boldsymbol{\theta}_D)}{\|\boldsymbol{\Phi}_{k,q}(\boldsymbol{\theta}_D)\|^2} \tag{65}$$

accounts for the arbitrary scaling of the mode shapes. The prediction-error parameters $c_{\lambda,r}$ and $c_{\Phi,r}$ are treated as fixed. Accordingly, the prediction-error parameters $\boldsymbol{\Theta}_\varepsilon$ of the general formulation in Section 2 are not treated as uncertain either. Thus, only the deterioration parameters $\boldsymbol{\Theta}_D$ are inferred in the present example.

The prediction errors associated with the matched modal quantities are assumed mutually independent within each monitoring stage, including the eigenvalue errors and the errors of the individual mode-shape coordinates, and independent among monitoring stages. Under these assumptions, the likelihood contribution of the monitoring data at stage $k$ is

$$L_k(\boldsymbol{\theta}_D) = \prod_{r=1}^{n_m} \left[ \mathcal{N}\left(\varepsilon_{\lambda,k,r}(\boldsymbol{\theta}_D); 0, \sigma_{\lambda,k,r}^2\right) \prod_{s=1}^{n_\Phi} \mathcal{N}\left(\varepsilon_{\Phi,k,r,s}(\boldsymbol{\theta}_D); 0, \sigma_{\Phi,k,r}^2\right) \right]. \tag{66}$$

Owing to the assumed independence among monitoring stages, the conditional likelihood introduced in Section 2.2 reduces in the present example to $L_{k|1:k-1}(\boldsymbol{\theta}_D) = L_k(\boldsymbol{\theta}_D)$ and consequently

$$L_{1:k}(\boldsymbol{\theta}_D) = \prod_{i=1}^{k} L_i(\boldsymbol{\theta}_D). \tag{67}$$

The present monitoring model is therefore a special case of the general formulation in Section 2, with fixed structural-behavior and prediction-error parameters and independent Gaussian prediction errors.

For its implementation within the Sequential BUS formulation of Section 3, the incremental likelihood multiplier $c_k$ at monitoring stage $k$ must satisfy $c_k\, L_k(\boldsymbol{\theta}_D) \leq 1$ for all $\boldsymbol{\theta}_D$. The computational efficiency of Sequential BUS with Subset Simulation depends on the choice of these stagewise multipliers. At a given monitoring stage, an unnecessarily small $c_k$, corresponding to an overly conservative upper bound on the incremental likelihood, reduces the probability of the updated BUS event relative to the preceding one and can increase the number of required Subset Simulation levels. Since the cumulative likelihood multiplier is obtained recursively as $c_{1:k} = c_{1:k-1}c_k$, conservatism in the individual $c_k$ values also accumulates in $c_{1:k}$, leading to an increasingly conservative representation of the cumulative BUS event. In BUS with Subset Simulation, such conservative scaling is undesirable because additional subset levels increase the dependence among the MCMC-generated conditional samples [39].

For the Gaussian likelihood in Eq. (66), an analytical upper bound can be obtained by setting all prediction errors to zero. In the present application, however, this bound is overly conservative because the model-predicted modal properties are restricted to the finite set of brace-damage states and a zero prediction error is generally not attainable. This discrete structure allows the maximum attainable likelihood to be determined directly for the monitoring data observed at each stage. The OpenSees model is deterministic conditional on the brace-damage state, such that a numerical modal analysis can be performed in advance for every $\boldsymbol{d} \in \mathcal{D}$, where $\mathcal{D}$ contains the $2^{n_b} = 2^{13} = 8192$ possible brace-damage states. The resulting modal properties are stored and subsequently used both for efficient likelihood evaluation during Sequential BUS and for determining the stage-specific likelihood multiplier.

Once the monitoring observation stage $k$ becomes available, the likelihood $L_k(\boldsymbol{d})$ is evaluated for every $\boldsymbol{d} \in \mathcal{D}$ by comparing the corresponding precomputed modal properties with the observed modal quantities. The maximum attainable likelihood for the observed data is therefore

$$L_{k,\max} = \max_{\boldsymbol{\theta}_D} L_k(\boldsymbol{\theta}_D) = \max_{\boldsymbol{d}\in\mathcal{D}} L_k(\boldsymbol{d}), \tag{68}$$

and the stage-specific likelihood multiplier is selected as

$$c_k = \frac{1}{L_{k,\max}}, \quad \text{or equivalently} \quad \ell_k = -\log c_k = \max_{\boldsymbol{d}\in\mathcal{D}} \log L_k(\boldsymbol{d}) \tag{69}$$

Thus, $c_k$ is determined separately after the monitoring data at stage $k$ have been observed, using an exhaustive search over the discrete structural states. This gives the largest admissible stage-specific multiplier while ensuring $c_k\, L_k(\boldsymbol{\theta}_D) \leq 1$ for all deterioration-parameter realizations. The synthetic monitoring observations used for this procedure are introduced in Section 4.4.

### 4.4 Synthetic monitoring scenario

A synthetic monitoring scenario is considered over a period of 20 years, with one monitoring stage in each year $t_k$, $k = 1, \dots, 20$. For years 1 - 19, the structure is assumed to remain in the undamaged state. At year 20, brace 6 is assumed to have failed, with the brace indices defined in Figure 1. This prescribed sequence of brace-level deterioration states defines the structural scenario used to generate the synthetic monitoring data.

At each monitoring stage, the corresponding brace configuration is represented in the OpenSees model and subjected to horizontal white-noise excitation at the supports. The resulting horizontal acceleration responses are simulated at the sensor locations shown in Figure 2. To represent sensor measurement error, the noise-free acceleration time series are contaminated with Gaussian white noise corresponding to a 2% root-mean-square noise-to-signal ratio. The covariance-driven stochastic subspace identification method (SSI-COV) [40] is subsequently applied to the noisy acceleration responses to identify the modal eigenvalues and mode shapes used as monitoring observations. The resulting identified modal frequencies and mode shapes for years 1 - 20 are shown in Figure 3. The failure of brace 6 is reflected particularly in distinct changes in the first and third modal frequencies and in the third mode shape. The modal eigenvalues corresponding to the modal frequencies in Figure 3 are used in the likelihood formulation of Section 4.3.

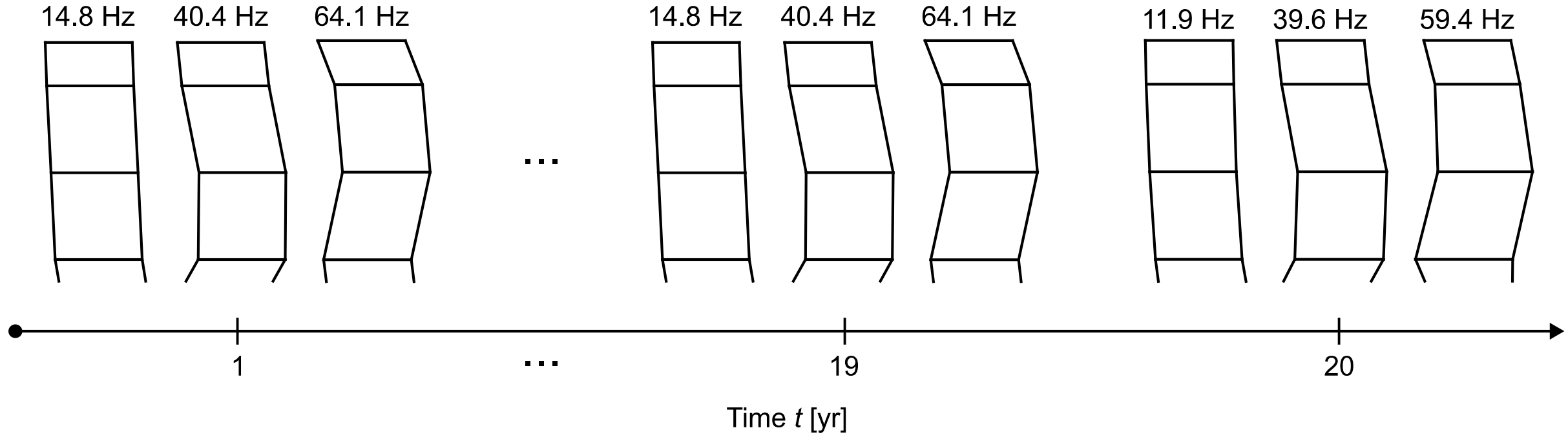

**Figure 3:** Synthetic monitoring observations comprising the identified modal frequencies and corresponding mode shapes. Years 1-19 correspond to the intact structure, while the year-20 observations reflect failure of brace 6.

## 4.5 Sequential inference

The monitoring data are assimilated sequentially using Sequential BUS with Subset Simulation introduced in Section 3. At each monitoring stage $k$, the resulting sample population represents the posterior distribution of the deterioration parameters conditional on all observations available up to that stage. The following results first illustrate the temporal evolution of selected marginal posterior distributions and subsequently examine the posterior distributions immediately before and after the change in the modal properties observed in year 20. Hotspots 10 and 18, indicated in Figure 1, are selected as representative examples.

The results presented below correspond to one realization of the Sequential BUS analysis. Repeated Sequential BUS runs produced closely comparable posterior estimates throughout the monitoring period, including after assimilation of the year-20 observation, with only small run-to-run variability. The final posterior obtained after year 20 was additionally verified against a conventional BUS analysis in which the complete monitoring dataset was assimilated in a single updating step; the resulting posterior distributions were closely consistent with those obtained from Sequential BUS.

Figure 4 shows the evolution of the filtered marginal distributions of $B_S$, $B_{SIF}$, $\ln C$, and $A_0$ for hotspots 10 and 18 in terms of their posterior means and 90% credible intervals. During the first 19 years, the monitoring data consist of modal observations corresponding to the intact structure. Over this period, the posterior means remain comparatively stable, while the marginal credible intervals show only a gradual contraction as monitoring information accumulates. This behavior is consistent with the information provided by the repeated intact observations: parameter combinations associated with deterioration trajectories that would have resulted in joint and brace failure before the respective monitoring time become increasingly inconsistent with the data, whereas the observations provide only limited discrimination among the remaining subcritical deterioration trajectories.

The observation of changed modal properties in year 20 produces a more pronounced and parameter-specific redistribution of posterior probability. The changes are comparatively modest for hotspot 10 but substantially stronger for hotspot 18. Depending on the parameter, the year-20 update results in shifts of the posterior mean, changes in the marginal uncertainty, or both. In particular, the marginal uncertainty does not necessarily decrease when the more informative year-20 observation is assimilated. This reflects the fact that the monitoring data constrain combinations of fatigue-model parameters through their common effect on the deterioration state rather than identifying each parameter independently.

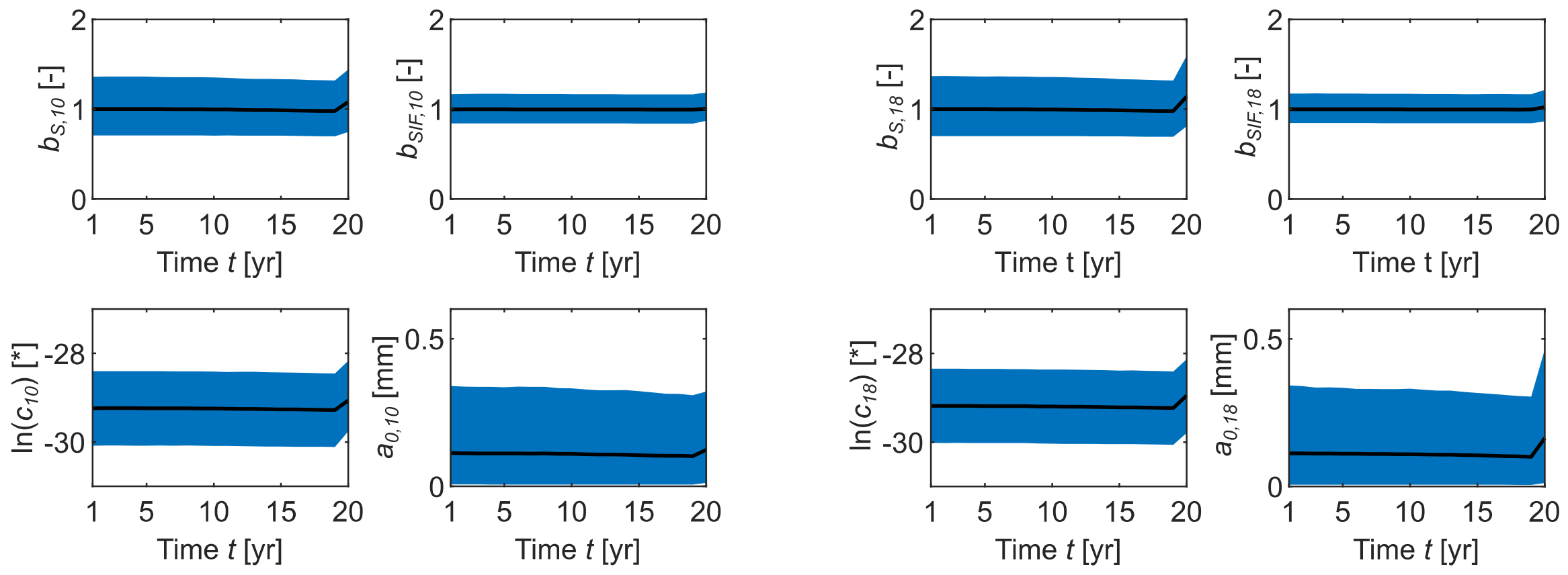

**Figure 4:** Evolution of the filtered marginal posterior distributions of $B_{S,j}$, $B_{\mathrm{SIF},j}$, $\ln C_j$, and $A_{0,j}$ for hotspots 10 (left) and 18 (right), represented by posterior means and 90% credible intervals.

Figure 5 to Figure 8 compare the marginal and bivariate posterior distributions of the deterioration-model parameters $B_S$, $B_{\mathrm{SIF}}$, $\ln C$, and $A_0$ for hotspots 10 and 18 in years 19 and 20 with their corresponding priors. In year 19, following repeated observations of modal properties corresponding to the intact structure, the posterior distributions remain close to the priors. Only moderate changes in location and spread are observed, while the bivariate posterior samples continue to occupy connected regions that broadly follow the prior dependence structure. This limited update is consistent with the information provided by the intact observations: deterioration trajectories that would have produced a brace failure before year 19 are progressively downweighted, whereas a broad range of subcritical deterioration histories remains compatible with the monitoring data.

The posterior distributions change more markedly after assimilation of the year-20 observation. The effect is parameter and hotspot dependent, with more pronounced redistribution for hotspot 18 than for hotspot 10. The marginal posterior distributions nevertheless remain broadly single-peaked, although some become more asymmetric or broader. The corresponding bivariate distributions exhibit localized concentrations of posterior samples, but these regions are closely spaced and remain part of a connected posterior domain rather than forming clearly separated modes. The year-20 monitoring information therefore constrains particular combinations of deterioration-model parameters without uniquely identifying the individual parameters. The resulting posterior structure reflects dependence and compensation among the fatigue-model parameters, while posterior means and 90% credible intervals remain meaningful summaries of the marginal distributions. The bivariate distributions provide additional information on the dependence structure and local concentration of posterior probability.

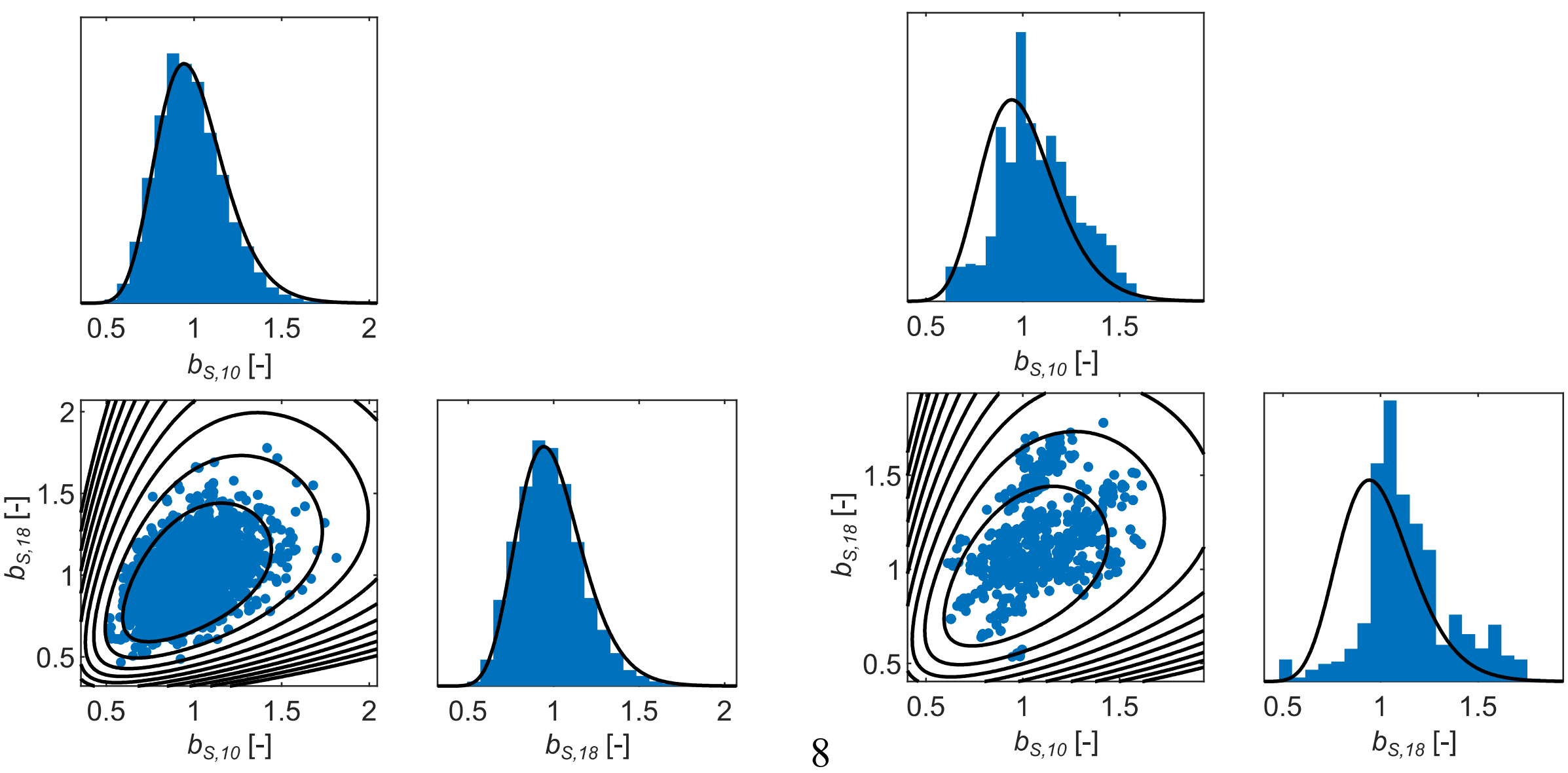

**Figure 5:** Prior and filtered marginal and bivariate posterior distributions of $B_{S,10}$ and $B_{S,18}$ in year 18 (left) and year 20 (right).

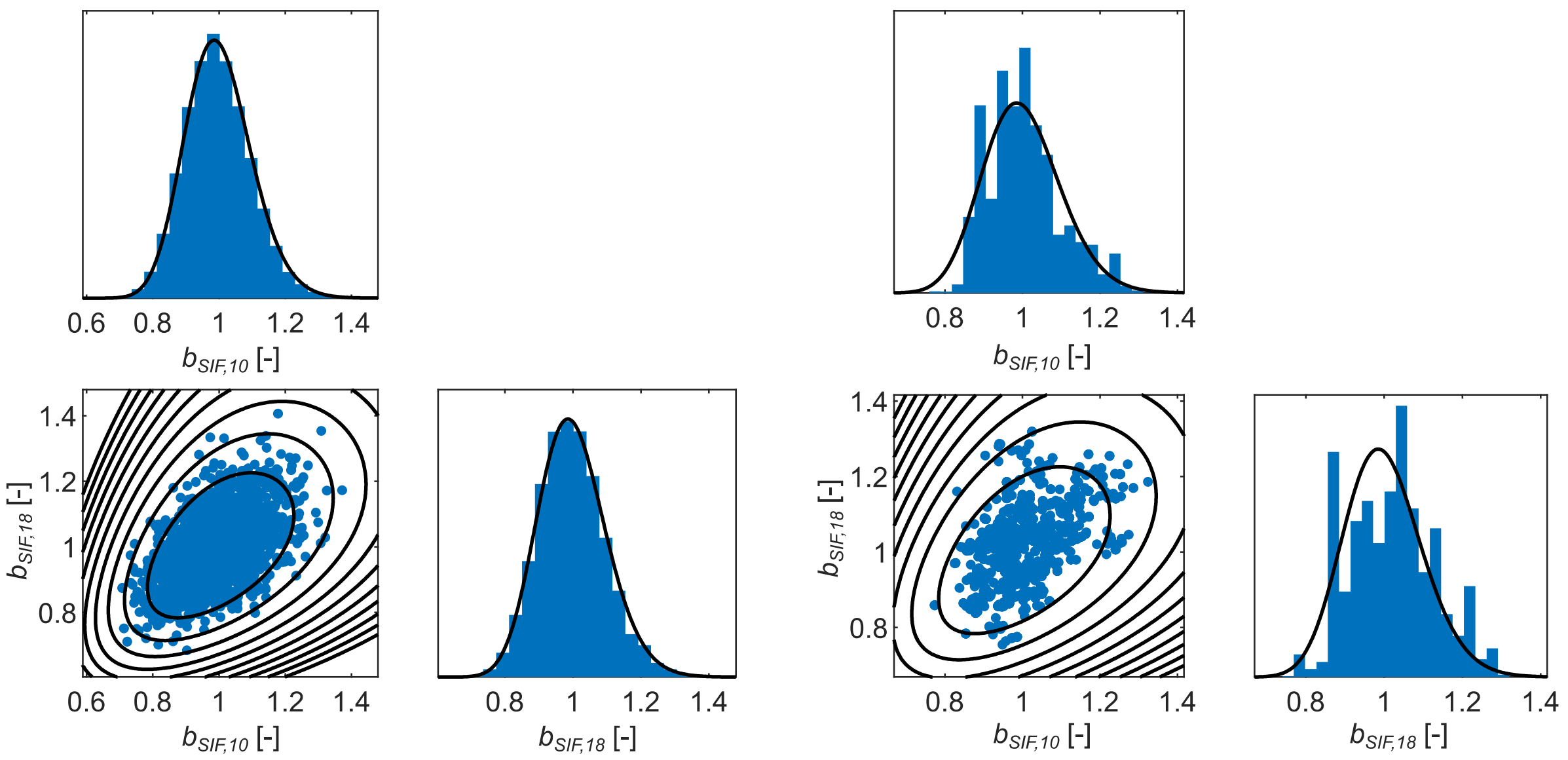

**Figure 6:** Prior and filtered marginal and bivariate posterior distributions of $B_{\mathrm{SIF},10}$ and $B_{\mathrm{SIF},18}$ in year 18 (left) and year 20 (right).

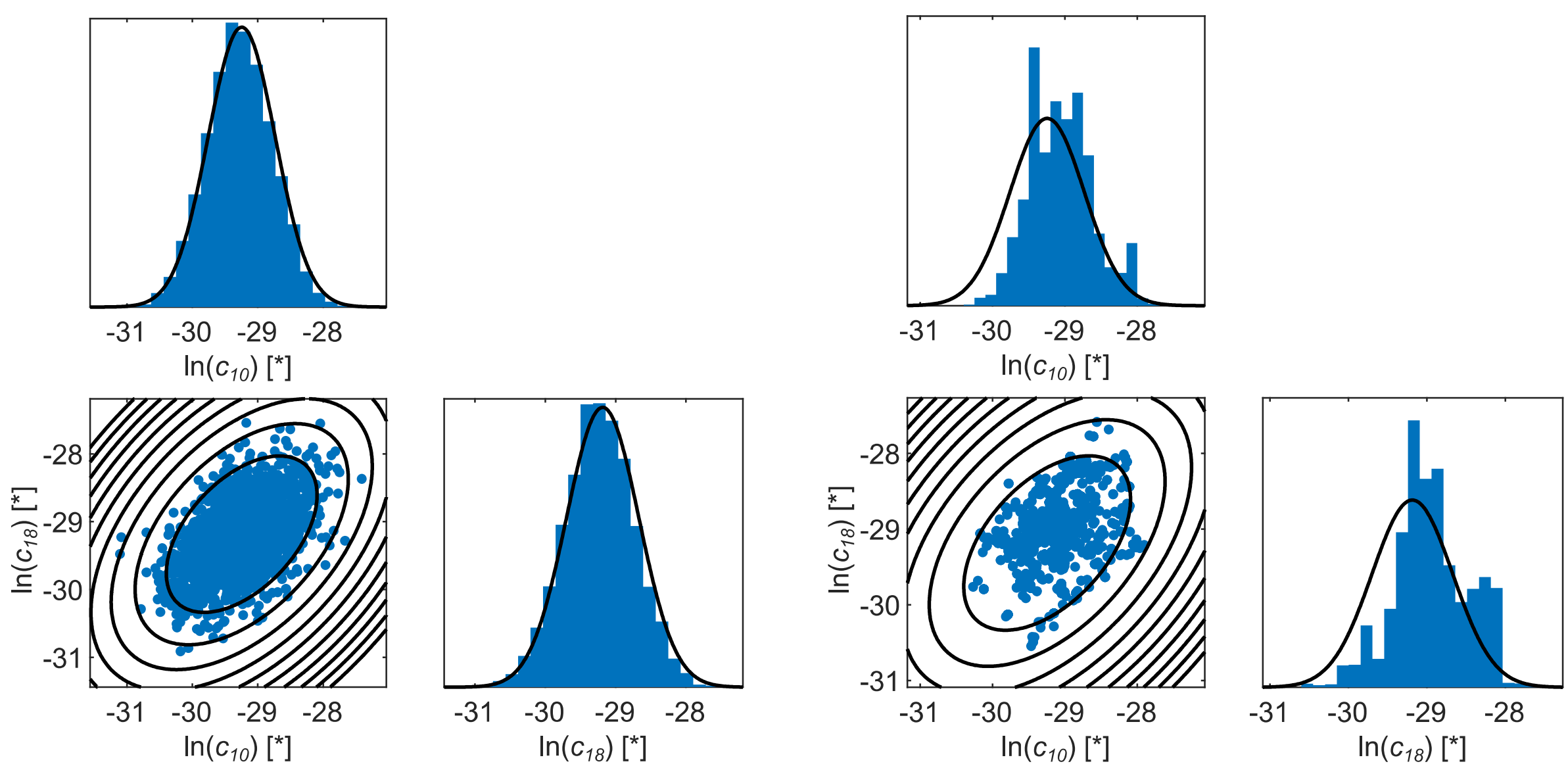

**Figure 7:** Prior and filtered marginal and bivariate posterior distributions of $\ln C_{10}$ and $\ln C_{18}$ in year 18 (left) and year 20 (right).

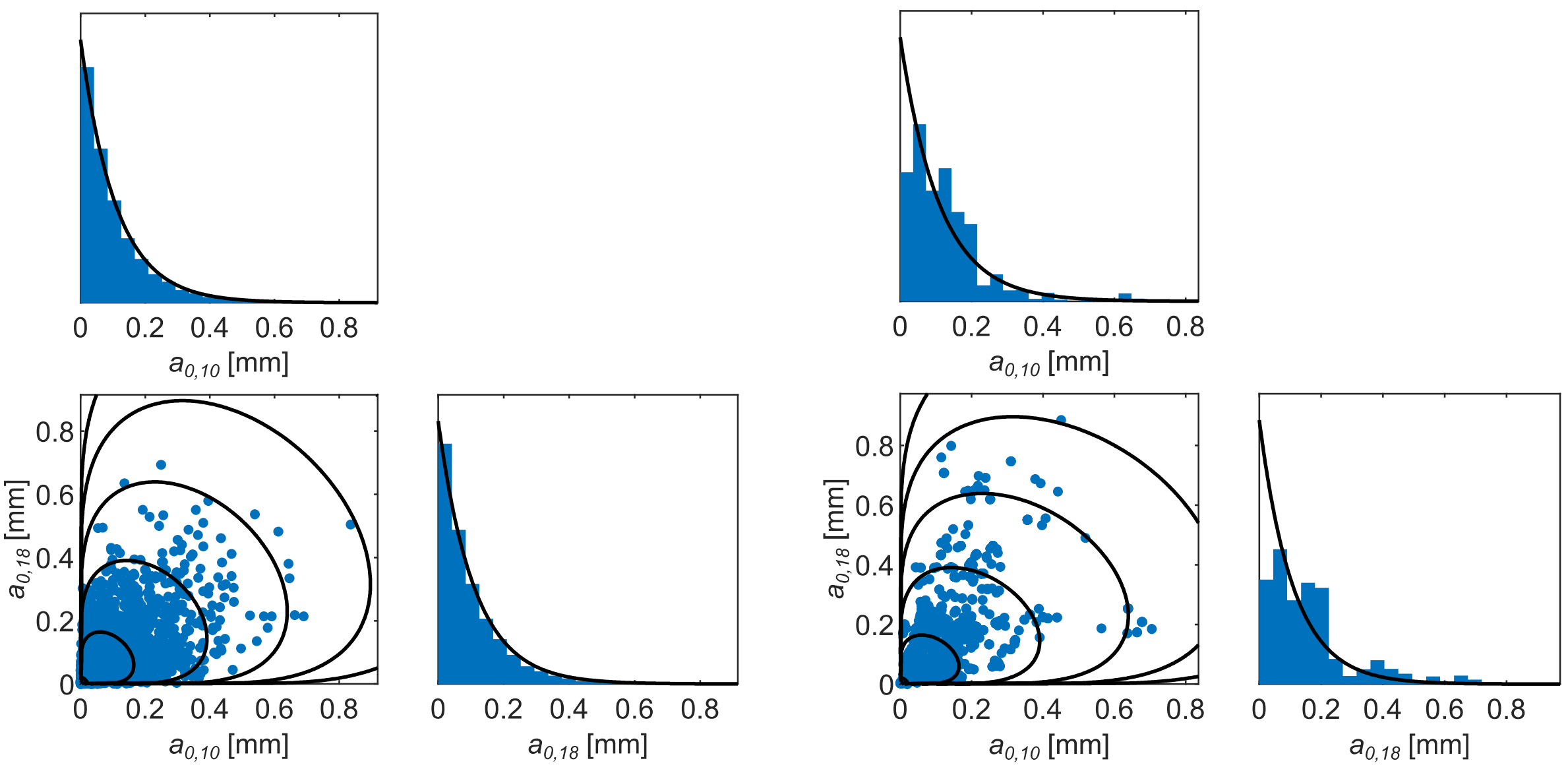

**Figure 8:** Prior and filtered marginal and bivariate posterior distributions of $A_{0,10}$ and $A_{0,18}$ in year 18 (left) and year 20 (right).

### 4.6 System-level diagnosis

Table 1 summarizes the system-level diagnosis obtained by propagating the posterior samples of the deterioration parameters through the system deterioration model. During the first 19 monitoring stages, the intact system state remains overwhelmingly dominant. Its posterior probability decreases only gradually and is 0.9945 in year 19. The small probability assigned to alternative states reflects deterioration trajectories that remain compatible with the observed modal properties but are substantially less probable under the deterioration model. The repeated observations corresponding to the intact structure therefore provide primarily survival-type information: deterioration histories that would have resulted in brace failure before the respective monitoring time are progressively downweighted, while a broad range of subcritical deterioration histories remains compatible with the observations.

The diagnosis changes markedly after assimilation of the changed modal properties observed in year 20. Posterior probability is concentrated on three deterioration states: failure of brace 6 only, corresponding to the true simulated state, with probability 0.51; failure of brace 5 only, with probability 0.48; and simultaneous failure of braces 5 and 6, with probability 0.01. The modal properties predicted for these states are very similar at the monitored degrees of freedom, such that the monitoring data alone provide only limited discrimination between them.

Importantly, the system-level diagnosis is not based solely on the agreement between the predicted and observed modal properties. The posterior probabilities also reflect the probability of the corresponding deterioration histories under the deterioration model. This is particularly evident for the simultaneous failure of braces 5 and 6: although this state provides the largest likelihood for the year-20 observation, its posterior probability remains small because simultaneous failure of both braces is substantially less probable under the deterioration model than a single-brace failure. The two single-brace failure states are therefore strongly favored, while their similar modal properties lead to substantial remaining ambiguity between braces 5 and 6. Since the fatigue hotspots associated with these braces have identical probabilistic properties, their nearly equal posterior probabilities primarily reflect the limited ability of the monitoring observations to distinguish the two structural states.

The result illustrates the probabilistic nature of the diagnosis: the monitoring data restrict the structural condition to a small set of plausible states, while the deterioration model weights these states according to the plausibility of the deterioration processes required to produce them. The remaining system-state uncertainty is retained in the posterior deterioration-model samples and propagated to the subsequent deterioration prognosis and reliability prediction.

**Table 1:** Posterior probabilities of the diagnosed system deterioration states at monitoring stage $k$, conditional on the monitoring data $\boldsymbol{z}_{1:k}$. The true system state is highlighted in green.

| Stage $k$ | System state 1 | System state 2 | System state 3 | System state 4 | System state 5 |
|---|---|---|---|---|---|
| 1 | 1.0 | – | – | – | – |
| 2 | 1.0 | – | – | – | – |
| 3 | 1.0 | – | – | – | – |

| Stage $k$ | System state 1 | System state 2 | System state 3 | System state 4 | System state 5 |
|---|---|---|---|---|---|
| 4 | 0.9999 | 0.0001 | – | – | – |
| 5 | 0.9998 | 0.0001 | 0.0001 | – | – |
| 6 | 0.9998 | 0.0001 | 0.0001 | – | – |
| 7 | 0.9997 | 0.0001 | 0.0001 | 0.0001 | |
| 8 | 0.9998 | 0.0002 | – | – | – |
| 9 | 0.9995 | 0.0001 | 0.0002 | 0.0002 | – |
| 10 | 0.9995 | 0.0001 | 0.0002 | 0.0002 | – |
| 11 | 0.9988 | 0.0005 | 0.0006 | 0.0001 | – |
| 12 | 0.9985 | 0.0008 | 0.0007 | – | – |
| 13 | 0.9981 | 0.0011 | 0.0005 | 0.0003 | – |
| 14 | 0.9973 | 0.0012 | 0.0009 | 0.0003 | – |
| 15 | 0.9974 | 0.0011 | 0.0010 | 0.0005 | – |
| 16 | 0.9970 | 0.0008 | 0.0013 | 0.0008 | 0.0001 |

| Stage $k$ | System state 1 | System state 2 | System state 3 | System state 4 | System state 5 |
|---|---|---|---|---|---|
| 17 | 0.9963 | 0.0009 | 0.0018 | 0.0010 | – |
| 18 | 0.9956 | 0.0011 | 0.0016 | 0.0015 | 0.0002 |
| 19 | 0.9945 | 0.0015 | 0.0022 | 0.0018 | – |
| 20 | 0.5122 | 0.4800 | 0.0078 | – | – |

## 4.7 Hotspot-level prognosis

Figure 9 to Figure 11 illustrate the propagation of deterioration uncertainty to hotspot-level crack-growth predictions. Each figure shows the pointwise 90% credible intervals of the crack-depth trajectories for hotspots 10 and 18 together with corresponding sample trajectories, which provide additional insight into the occurrence of rapidly accelerating crack growth and fatigue failure.

Figure 9 shows the prior crack-growth predictions. The different fatigue characteristics of the two hotspots are clearly reflected in both the credible intervals and the individual trajectories. Crack growth at hotspot 10, with a design fatigue life of 40 years, is generally comparatively slow, although some prior realizations exhibit accelerated growth and reach the critical crack depth within the considered time horizon. Hotspot 18, with a design fatigue life of 25 years, is substantially more fatigue-critical: its prior uncertainty increases notably with time and a larger fraction of the sample trajectories reaches the critical crack depth before year 25.

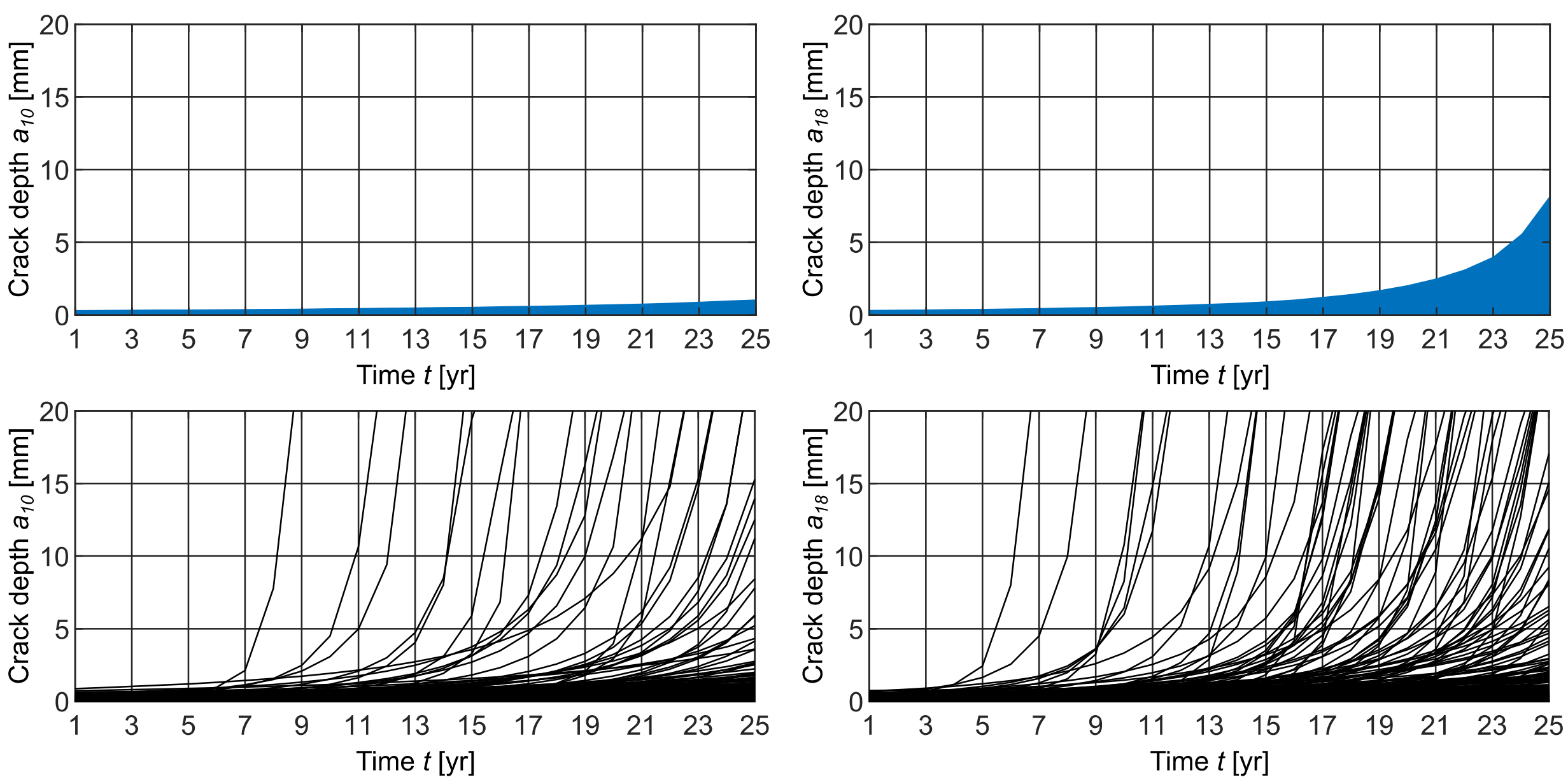

**Figure 9:** Prior 90% credible intervals of the crack-depth trajectories for hotspots 10 (upper left) and 18 (upper right), with corresponding sample trajectories shown in the lower panels.

Figure 10 shows the corresponding posterior predictions conditional on monitoring data up to year 19. The repeated observations of modal properties corresponding to the intact structure provide survival-type information by excluding or strongly downweighting deterioration trajectories that would have caused a joint and brace failure before the current monitoring time. The resulting crack-growth distributions are therefore considerably more confined up to year 19, particularly for hotspot 18. The individual trajectories nevertheless show that rapid crack growth shortly after year 19 remains possible. Thus, the monitoring information indicates that critical deterioration has effectively not occurred by the current monitoring time, while substantial uncertainty remains in the subsequent crack-growth evolution and time to failure.

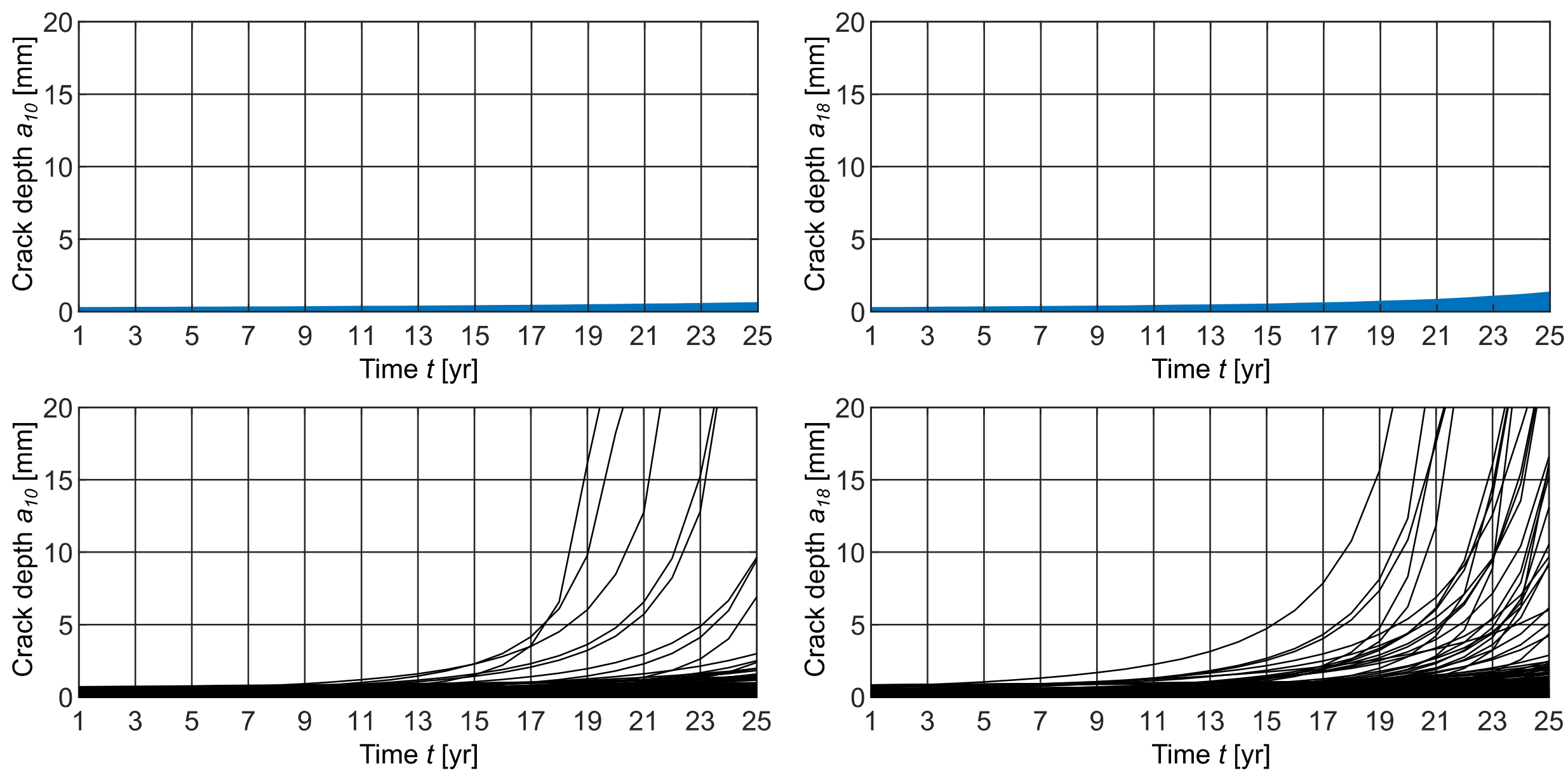

**Figure 10:** Posterior 90% credible intervals of the crack-depth trajectories for hotspots 10 (upper left) and 18 (upper right), conditional on monitoring data up to year 19, with corresponding sample trajectories shown in the lower panels.

Figure 11 shows the predictions after incorporating the observation of changed modal properties in year 20. The update has markedly different effects on the two hotspots. For hotspot 18, posterior probability shifts strongly toward deterioration trajectories that remain subcritical up to year 19 but subsequently undergo rapid crack growth. A substantial fraction of the posterior trajectories reaches the critical crack depth during year 20, while others fail only at later times or remain subcritical over the considered horizon. This spread reflects the ambiguity of the monitoring observation: the changed modal properties indicate that structural deterioration has occurred, but they do not uniquely identify which of the four candidate hotspots associated with the affected braces has reached failure. Consequently, failure of hotspot 18 becomes substantially more plausible without being uniquely implied by the monitoring data.

For hotspot 10, the survival-type information accumulated up to year 19 continues to suppress trajectories associated with earlier failure. Beyond year 20, however, rapidly growing trajectories remain plausible, and the 90% credible interval becomes slightly wider than that obtained when conditioning only on data up to year 19. This is consistent with the statistical dependence between the hotspot deterioration processes: observing structural response indicative of a failure at another location also increases the plausibility of more severe deterioration trajectories at hotspot 10. The monitoring information therefore redistributes probability among the possible local deterioration histories according to their compatibility with the observed structural response, while substantial uncertainty remains regarding both the location of the observed failure and the subsequent evolution of deterioration at the remaining hotspots.

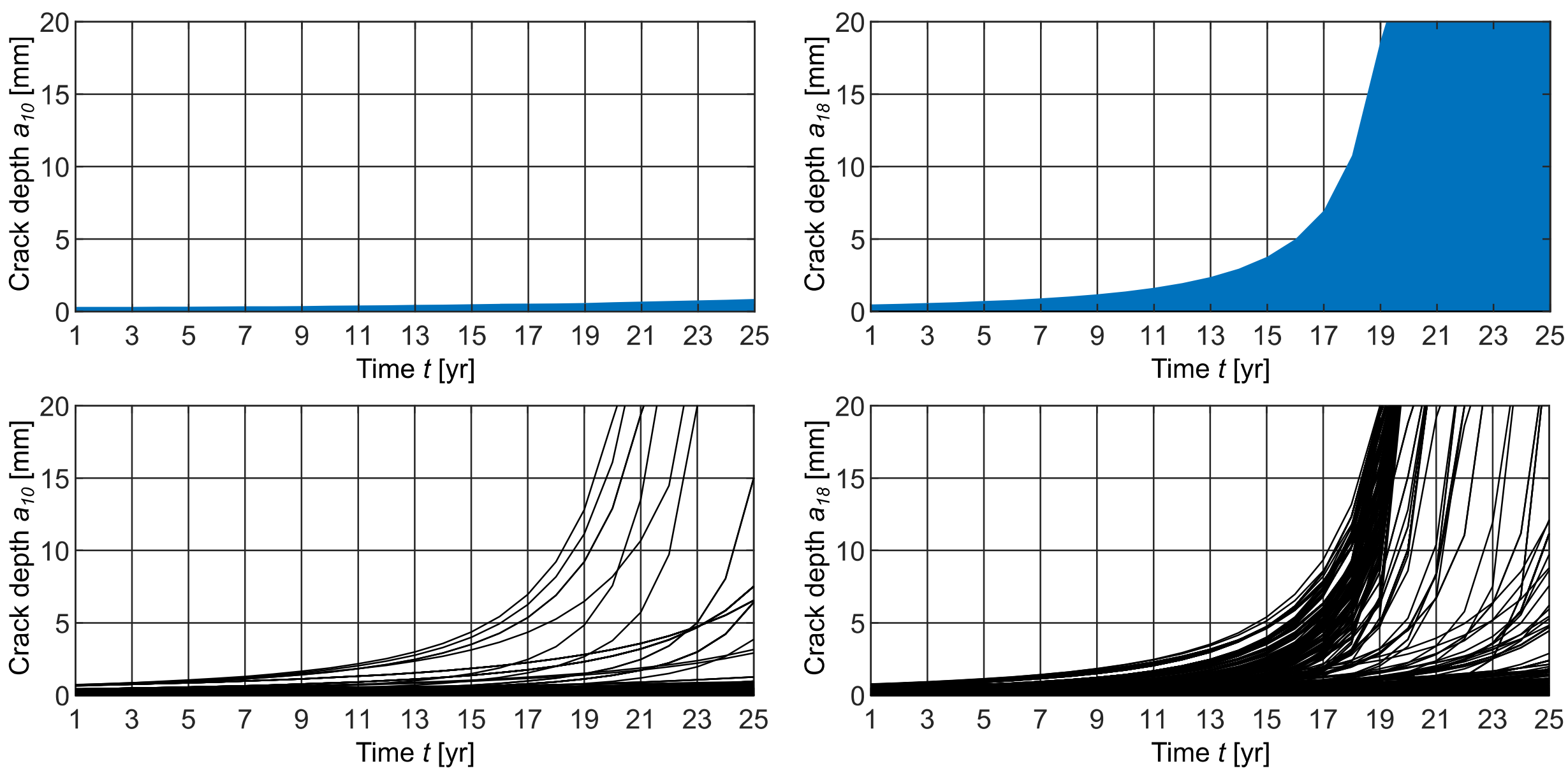

**Figure 11:** Posterior 90% credible intervals of the crack-depth trajectories for hotspots 10 (upper left) and 18 (upper right), conditional on monitoring data up to year 20, with corresponding sample trajectories shown in the lower panels.

### 4.8 Monitoring-informed reliability prediction

System reliability is evaluated following the nested reliability approach of Schneider et al. [14], with the revised formulation given in Schneider [27]. The structural performance model introduced in Section 2.3 is implemented in USFOS [33] as a nonlinear static pushover analysis. For the fixed horizontal load configuration shown in Figure 1, the applied load is increased incrementally until the ultimate load-carrying capacity of the deteriorated structure is reached. The performance model therefore maps the system deterioration state $\boldsymbol{D}(t)$ to the corresponding resistance

$$R(t) = h_P(\boldsymbol{D}(t)), \tag{70}$$

where $R(t)$ denotes the maximum magnitude of the prescribed horizontal load pattern that the structure can sustain at time $t$.

The USFOS model is treated as deterministic conditional on the deterioration state [14, 27]. For each realization $\boldsymbol{d}$, braces in the failed state are removed from the structural model and the prescribed horizontal load is progressively increased until structural collapse. The resulting ultimate load defines $h_P(\boldsymbol{d})$. Since the frame contains 13 braces, the resistances corresponding to all $2^{13} = 8192$ possible system deterioration states are precomputed and stored for use in the reliability analyses. The structural performance model is distinct from the linear-elastic OpenSees model used in Section 4.3: the latter provides the observation link from deterioration to the measured modal properties, whereas $h_P$ provides the consequence link from deterioration to structural resistance.

According to the deterioration model introduced in Section 4.2, $\boldsymbol{D}(t) = \boldsymbol{h}_D(t; \boldsymbol{\Theta}_D)$, such that the resistance can equivalently be expressed in terms of the deterioration parameters as

$$R(t) = h_P[\boldsymbol{h}_D(t; \boldsymbol{\Theta}_D)]. \tag{71}$$

Consider yearly intervals $(t_{n-1}, t_n]$, $n = 1, \dots, N_T$, and let $F_n^*$ denote system failure during interval $n$. Let $S_{\max}$ denote the annual maximum magnitude of the horizontal environmental load corresponding to the load configuration shown in Figure 1. Conditional on a realization $\boldsymbol{\Theta}_D = \boldsymbol{\theta}_D$, the deterioration state and hence the resistance at $t_n$ are deterministic. The corresponding conditional interval failure probability is

$$\Pr(F_n^* | \boldsymbol{\Theta}_D = \boldsymbol{\theta}_D) = 1 - F_{S_{\max}}(h_P[\boldsymbol{h}_D(t; \boldsymbol{\Theta}_D)]), \tag{72}$$

where $F_{S_{\max}}$ denotes the distribution function of the annual maximum horizontal load.

The cumulative failure event up to time $t_n$ is

$$F(t_n) = \bigcup_{s=1}^{n} F_s^*. \tag{73}$$

Assuming statistical independence of the annual maximum loads, the cumulative system failure probability conditional on the deterioration parameters becomes

$$p_F(\boldsymbol{\theta}_D, t_n) = \Pr[F(t_n) | \boldsymbol{\Theta}_D = \boldsymbol{\theta}_D] = 1 - \prod_{s=1}^{n} [1 - \Pr(F_n^* | \boldsymbol{\Theta}_D = \boldsymbol{\theta}_D)]. \tag{74}$$

The dependence of the interval failure probabilities induced by the common deterioration parameters and the evolving deterioration state is thereby retained through $\boldsymbol{\theta}_D$.

At monitoring stage $k$, the cumulative monitoring information is represented by the BUS event

$$\mathcal{O}_{1:k} = \{\Pi \leq c_{1:k}\, L_{1:k}(\boldsymbol{\Theta}_D)\}. \tag{75}$$

Following the nested reliability formulation, the joint event of cumulative system failure and the monitoring information can be expressed in the augmented BUS space as

$$F(t_n) \cap \mathcal{O}_{1:k} = \{\Pi \leq p_F(\boldsymbol{\Theta}_D, t_n)\, c_{1:k}\, L_{1:k}(\boldsymbol{\Theta}_D)\}. \tag{76}$$

The corresponding limit-state function is

$$g_{F\cap\mathcal{O}}(\boldsymbol{\theta}_D, \pi, t_n, k) = \pi - p_F(\boldsymbol{\theta}_D, t_n)\, c_{1:k}\, L_{1:k}(\boldsymbol{\theta}_D). \tag{77}$$

such that

$$F(t_n) \cap \mathcal{O}_{1:k} = \{g_{\mathrm{F}\cap\mathcal{O}}(\boldsymbol{\theta}_D, \pi, t_n, k) \leq 0\}. \tag{78}$$

Since $0 \leq p_F(\boldsymbol{\theta}_D, t_n) \leq 1$, the joint failure-and-observation event is nested within $\mathcal{O}_{1:k}$.

The monitoring-informed cumulative system failure probability is consequently [14, 27]

$$p_F(t_n|\boldsymbol{z}_{1:k}) = \Pr[F(t_n)|\boldsymbol{z}_{1:k}] = \Pr[F(t_n)|\mathcal{O}_{1:k}] = \frac{\Pr[F(t_n) \cap \mathcal{O}_{1:k}]}{\Pr(\mathcal{O}_{1:k})}. \tag{79}$$

The separate indices $k$ and $n$ distinguish the monitoring stage from the reliability evaluation time. The formulation therefore encompasses retrospective reliability assessment for $t_n < t_k$ (smoothing), current reliability assessment for $t_n = t_k$ (filtering), and monitoring-informed reliability prediction for $t_n > t_k$ (prediction).

The probability in Eq. (79) is evaluated using the two-stage BUS/Subset Simulation procedure described in Section 1 [26, 28]. In the first stage, Sequential BUS provides an augmented sample population conditional on $O_{1:k}$, whose marginal distribution in $\Theta_D$represents the posterior given the monitoring data. In the second stage, this population is used directly to initialize Subset Simulation toward the joint failure-and-observation event $F(t_n) \cap O_{1:k}$, rather than initiating a new simulation from the prior distribution. Consistent with the conditional Subset Simulation formulations in Straub et al. [26] and Jerez et al. [28] and the implementation in Section 3.4, $O_{1:k}$ is retained as a fixed conditioning domain throughout the intermediate Subset Simulation levels. Adaptive thresholds are applied to $g_{F \cap O}$, and at the final threshold the sampled domain becomes $F(t_n) \cap O_{1:k}$.

The same two-stage procedure is used to evaluate monitoring-informed failure probabilities of individual fatigue hotspots. For hotspot $j$, failure at time $t_n$ is defined by

$$g_{F_j}(\boldsymbol{\theta}_D, t_n) = a_{c,j} - a_j(t_n; \boldsymbol{\theta}_D), \tag{80}$$

with

$$F_j(t_n) = \{g_{F_j}(\boldsymbol{\theta}_D, t_n) \leq 0\} = \{A_j(t_n) \geq a_{c,j}\}. \tag{81}$$

The corresponding monitoring-informed hotspot failure probability is

$$p_{F,j}(t_n|\boldsymbol{z}_{1:k}) = \Pr[F_j(t_n)|\boldsymbol{z}_{1:k}]. \tag{82}$$

Since crack growth is monotonic in the adopted deterioration model, this quantity represents the cumulative probability that hotspot $j$ has failed by $t_n$ [31]. For $t_n = t_k$, it gives the current monitoring-informed failure probability, whereas for $t_n > t_k$ it provides the corresponding prognosis. For its evaluation by Subset Simulation, the augmented samples conditional on $O_{1:k}$ are used as the initial population, $O_{1:k}$ is retained as the fixed conditioning domain, and the intermediate thresholds are applied to $g_{F_j}$ until the failure event $F_j(t_n)$ is reached.

The resulting monitoring-informed failure probabilities are shown in Figure 12 to Figure 15. Figure 12 compares the prior cumulative fatigue failure probabilities of hotspots 10 and 18 with the corresponding posterior probabilities conditional on the monitoring data available up to year 19. At $t = t_{19}$, the posterior values represent filtered failure probabilities, whereas the curves for $t > t_{19}$ represent monitoring-informed predictions. Conditioning on the repeated observations of modal properties corresponding to the intact structure effectively excludes fatigue failure of either hotspot up to the current monitoring time, and the filtered failure probabilities at year 19 are therefore essentially zero. For future times, however, the posterior failure probabilities increase rapidly. Although they remain below the corresponding prior probabilities, the reduction is comparatively moderate. This is consistent with the crack-growth predictions in Section 4.7: the monitoring data strongly constrain the deterioration history up to year 19, but provide substantially less information about the subsequent crack-growth evolution. Consequently, trajectories that are compatible with survival up to the current monitoring time can still undergo rapid crack growth and lead to fatigue failure shortly thereafter.

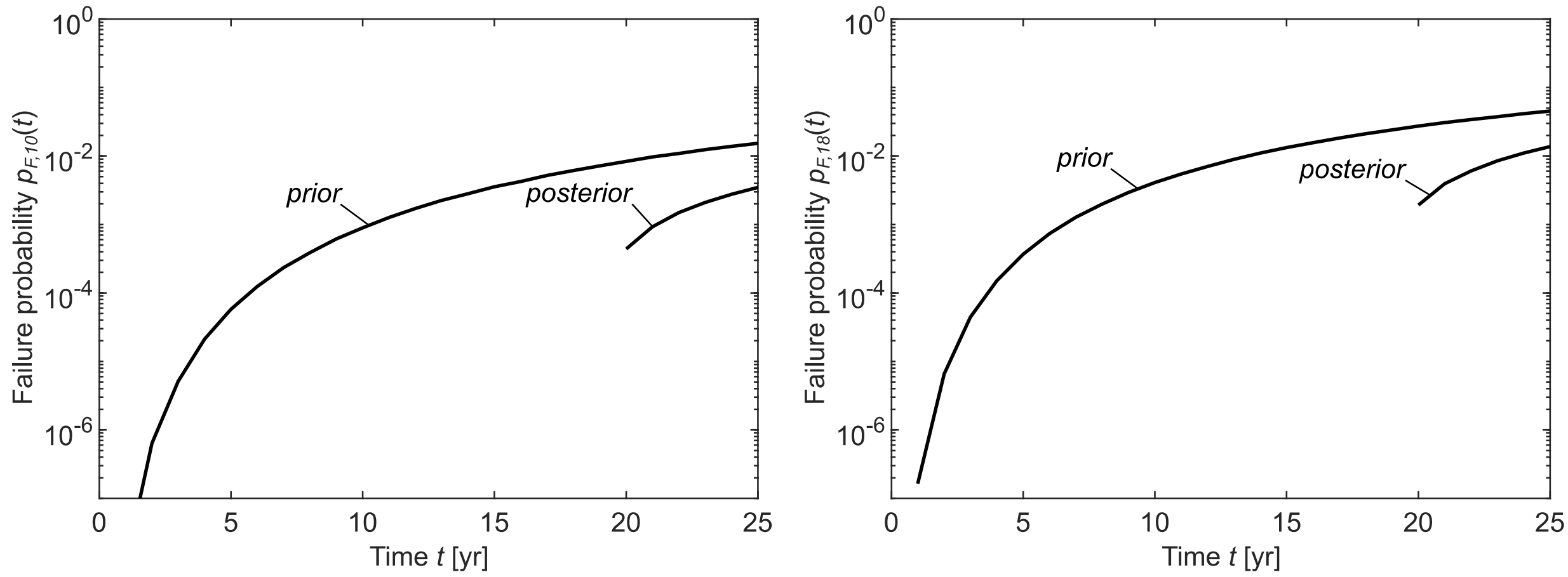

**Figure 12:** Prior and monitoring-informed cumulative fatigue failure probabilities of hotspots 10 (left) and 19 (right), conditional on monitoring data up to year 19.

Figure 13 shows the cumulative system failure probability obtained by sequential filtering over the monitoring period up to year 19, followed by prediction for later times. For each $t \leq t_{19}$, the filtered failure probability is conditional on the monitoring data available up to time $t$, whereas for $t > t_{19}$ the prediction is conditional on the data available up to year 19. Throughout this period, the repeated observations of modal properties corresponding to the intact structure assign a posterior probability greater than 0.99 to the undamaged system state. Accordingly, deterioration histories involving fatigue-induced brace failure are strongly downweighted, and the assessed system failure probability is substantially reduced relative to the prior prediction. The remaining system failure probability during the monitoring period primarily reflects the possibility of failure of the otherwise intact structure under environmental loading. For $t > t_{19}$, the curve represents a prognosis conditional on the monitoring information available up to year 19. The failure probability then increases again as future fatigue deterioration and brace failures become possible. This behavior is consistent with the crack-growth predictions in Section 4.7: the monitoring data strongly constrain the deterioration history up to the current monitoring time, while considerable uncertainty remains regarding the subsequent evolution of deterioration.

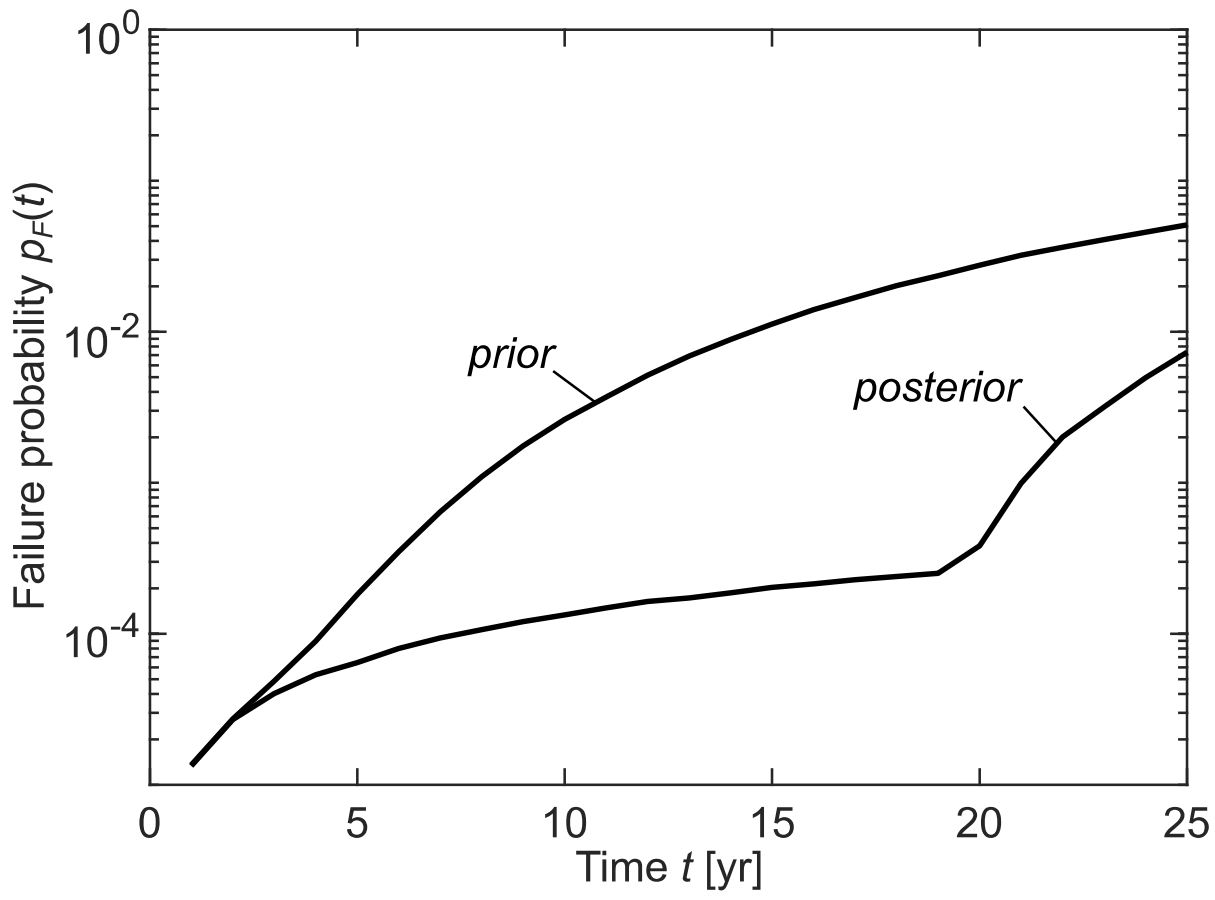

**Figure 13:** Prior and monitoring-informed cumulative system failure probabilities, conditional on monitoring data up to year 19.

Figure 14 shows the corresponding hotspot-level cumulative fatigue failure probabilities after incorporating the monitoring data from year 20, which contain an observation of changed modal properties. The update has markedly different effects on the two considered hotspots. For hotspot 18, the posterior cumulative failure probability increases sharply at year 20 to approximately 0.27. The same failure probability is obtained for the other three candidate hotspots associated with braces 5 and 6, reflecting the ambiguity of the monitoring observation regarding the location of the deterioration responsible for the observed change in modal properties. Thus, the monitoring data provide strong evidence that a fatigue-induced failure has occurred within this group of hotspots, but do not uniquely identify which hotspot has failed. For hotspot 10, in contrast, the filtered failure probability at year 20 remains essentially zero, indicating that failure of this hotspot is incompatible with the observed deterioration state at the current monitoring time. For $t > t_{20}$, however, its monitoring-informed failure probability increases and is higher than the corresponding prediction conditional only on data up to year 19. This is consistent with the crack-growth trajectories in Section 4.7: the observation of structural deterioration in year 20 increases the plausibility of more severe subsequent deterioration at hotspot 10 through the statistical dependence between the hotspot deterioration processes, even though hotspot 10 itself has not failed by year 20.

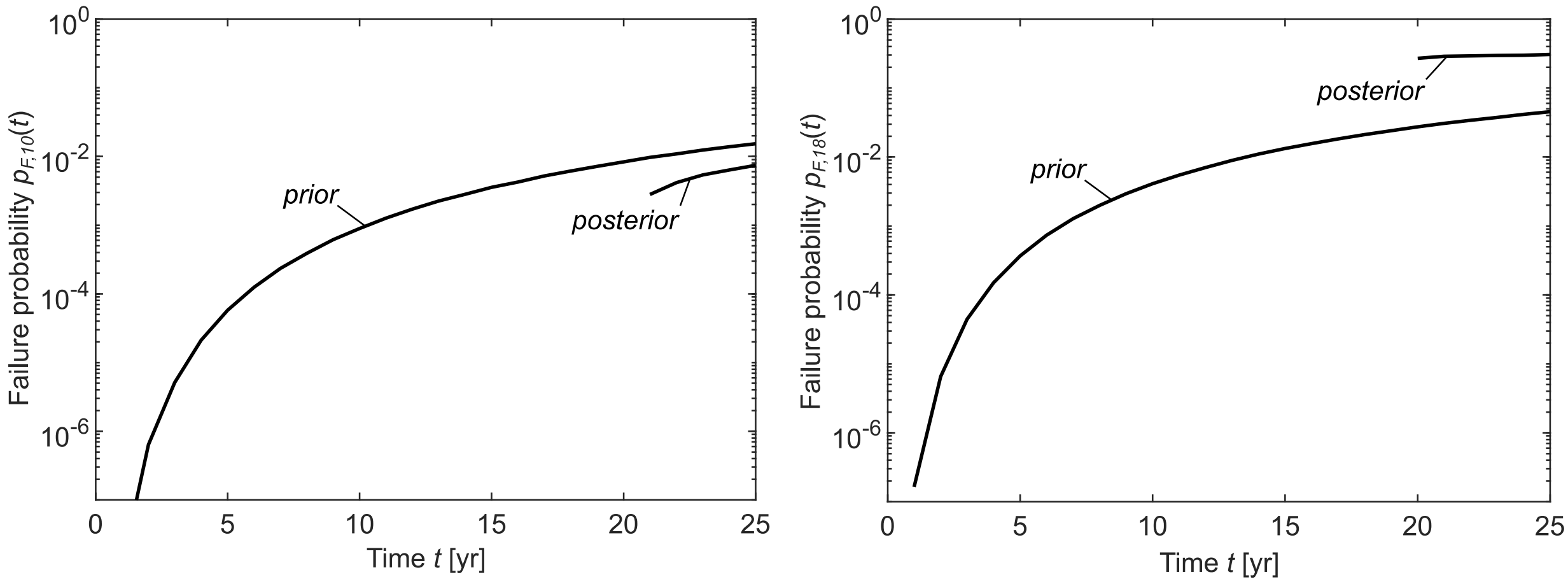

**Figure 14:** Prior and monitoring-informed cumulative fatigue failure probabilities of hotspots 10 (left) and 19 (right), conditional on monitoring data up to year 20.

Figure 15 shows the cumulative system failure probability obtained by sequential filtering over the monitoring period up to year 20, followed by prediction for later times. For each $t \leq t_{20}$, the filtered failure probability is conditional on the monitoring data available up to time $t$, whereas for $t > t_{20}$ the prediction is conditional on the data available up to year 20. Up to year 19, the monitoring-informed curve again represents successive filtered estimates associated with an intact structural state, and the system failure probability remains strongly reduced relative to the prior. At year 20, the observation of changed modal properties shifts substantial posterior probability to a damaged structural state, leading to a pronounced increase in the filtered system failure probability. For $t > t_{20}$, the curve represents a prognosis conditional on the monitoring information available up to year 20 and continues to increase as further deterioration of the already damaged structure becomes possible. Despite this increase, the system failure probability remains below the corresponding fatigue failure probability of hotspot 18. This reflects the structural redundancy of the system: failure of a fatigue hotspot and the associated loss of a brace changes the deterioration state but does not necessarily result in system failure. The structural consequences of the inferred deterioration are instead governed by the nonlinear system performance model.

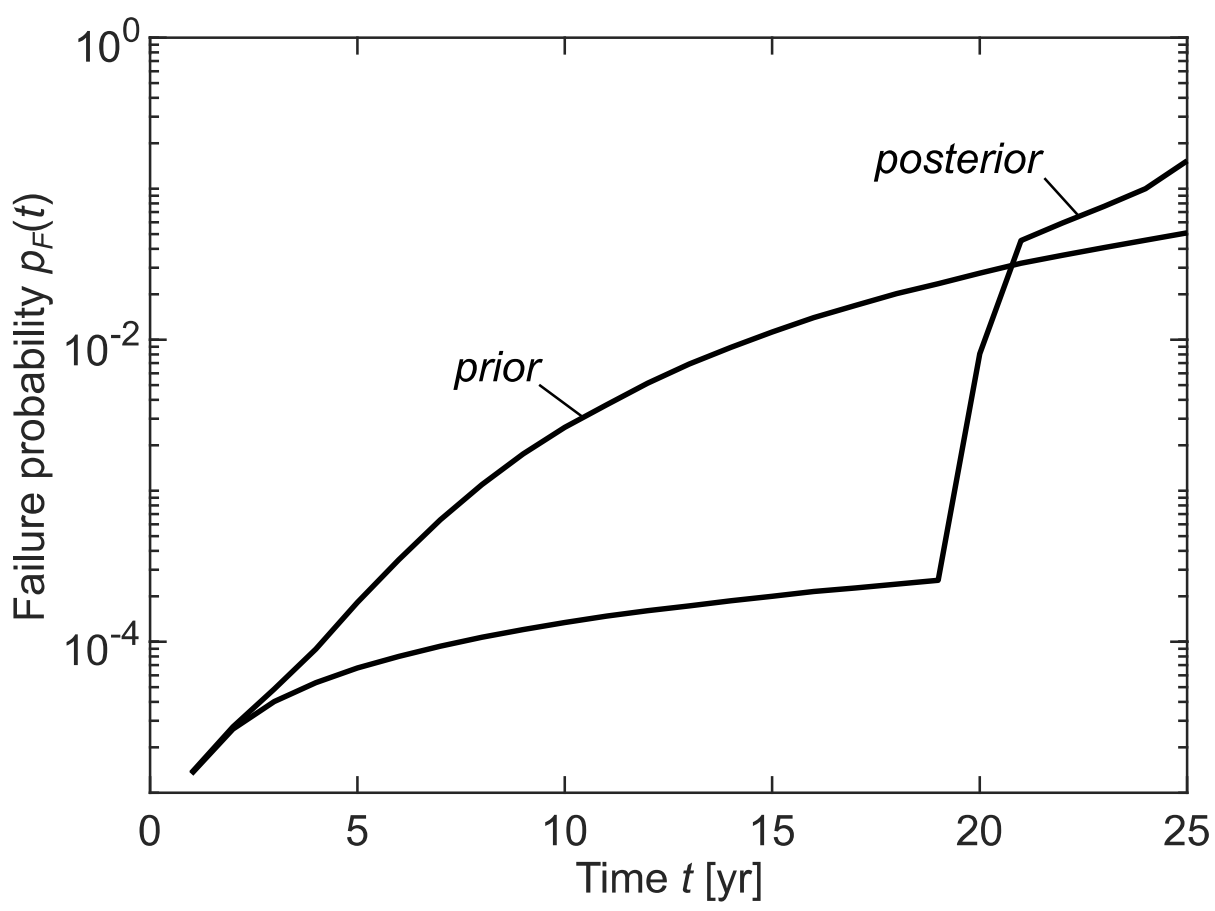

**Figure 15:** Prior and monitoring-informed cumulative system failure probabilities, conditional on monitoring data up to year 20.

# 5 Discussion

**Conceptual contribution and dependence on suitable models.** The proposed framework formalizes how response-based monitoring information is transferred to deterioration diagnosis, prognosis, and, where required, structural reliability prediction. Monitoring observations are interpreted through the structural behavior and prediction-error models to infer the uncertain parameters governing the latent deterioration process. The resulting posterior deterioration model then provides the common probabilistic basis for evaluating the current deterioration state and predicting its future evolution and associated structural consequences. An important aspect of this formulation is therefore the explicit separation between quantities inferred from the monitoring observations and quantities subsequently obtained through posterior prediction. Statistical dependence between local deterioration processes is retained in this propagation, such that monitoring information affecting one part of the deterioration process may also modify predictions at other locations.

A prerequisite for this model-based approach is the availability and adequate representation of the models forming this inference and prediction chain. First, the deterioration model must describe the relevant physical deterioration processes, their evolution, and, where relevant, their statistical dependence. Suitable models are not available with equal maturity for all deterioration mechanisms and may require substantial experimental and modelling effort [41]. Current work on hydrogen-induced stress corrosion cracking in quenched-and-tempered prestressing steel wires, for example, develops a fracture-mechanics-based crack-growth model using experimentally characterized defects and SHM-derived fatigue loading, with the longer-term objective of residual-life and probabilistic reliability assessment [42]. Second, the structural behavior model must provide a credible mapping from deterioration to the monitored response. In practical applications, this will generally require establishing and calibrating a probabilistic reference-state model through structural system identification, supported by measurements, prior engineering information, and, where relevant, explicit representation of environmental and operational variability [5, 6]. Where structural reliability prediction is pursued, suitable performance and load models are additionally required to relate the inferred deterioration state to structural failure; their formulation is well established within the structural reliability literature and is not a focus of the present discussion. The applicability of the proposed framework therefore depends on the availability and adequacy of these constituent models, while allowing advances in their development and identification to be incorporated within the same probabilistic architecture.

**Computational efficiency and robustness of Sequential BUS/Subset Simulation.** The Sequential BUS formulation provides a direct computational implementation of sequential Bayesian inference for successively growing monitoring datasets. By constructing the BUS events associated with consecutive monitoring stages as a nested sequence, the augmented conditional sample population obtained at one stage can be carried forward and further conditioned when new observations become available. Subset Simulation can therefore continue from the preceding posterior population instead of restarting from the original prior distribution at every monitoring stage. The nested construction is not required for the Bayesian validity of the individual BUS representations, but specifically enables this direct continuation of Subset Simulation and reuse of the information accumulated during the preceding stages.

This computational advantage does not make the method fully recursive in the sense of Bayesian filtering. At monitoring stage $k$, the BUS event is defined through the cumulative likelihood $L_{1:k}$. Although this likelihood can formally be factorized as $L_{1:k} = L_{1:k-1}L_{k|1:k-1}$, evaluations of historical likelihood contributions are generally still required for newly proposed parameter realizations generated during conditional sampling. The computational cost may therefore increase as the monitoring history grows. A further important determinant of efficiency is the choice of the BUS likelihood multiplier. An unnecessarily conservative likelihood bound reduces the probability of the corresponding BUS event and increases the number of conditional levels required to reach it. In the present example, the likelihood associated with an individual monitoring observation depends on a finite set of discrete structural damage states. The maximum incremental likelihood can therefore be determined exactly for the observed data by evaluating these states, providing the tight stagewise multiplier

$$c_k = [\max_{\boldsymbol{d} \in \mathcal{D}} L_{k|1:k-1}(\boldsymbol{d})]^{-1}.$$

This yields the most efficient admissible stagewise BUS representation while preserving the nested-event construction. For applications in which such an exact likelihood bound is unavailable, analytical bounds or adaptive BUS strategies [39] provide alternatives. The cumulative multiplier resulting from the sequential construction may nevertheless be more conservative than a multiplier obtained directly for the cumulative likelihood; where subsequent reliability analyses are performed from a fixed posterior, the BUS representation can therefore be rescaled without changing the underlying posterior distribution.

Sequential BUS also inherits the dependence of Subset Simulation on adequate sampling from the intermediate conditional distributions. Difficulties of standard Subset Simulation have been reported for problems involving separated important regions or complex changes in the geometry of the intermediate domains [43]. Recent work by Sharma et al. [44], however, provides a useful clarification of these observations: the difficulty is attributed primarily to insufficient exploration by the MCMC sampler used within Subset Simulation rather than to the conditional-probability decomposition itself. Sharma et al. demonstrate that conditional samplers combining local moves with explicit global exploratory transitions can substantially improve the exploration of separated and multimodal conditional regions. This suggests that robustness in such problems can be improved within the existing Subset Simulation framework by replacing or augmenting the conditional sampler, without changing the underlying Sequential BUS formulation.

The conditional sampler employed here follows the adaptive scheme of Papaioannou et al. [45] and provides efficient local exploration in standard normal space. For applications in which strongly separated conditional regions are anticipated, samplers incorporating additional global exploration may offer improved robustness [44]. As with MCMC methods generally, finite-sample detection of all relevant regions cannot be guaranteed, so repeated runs and problem-specific diagnostics remain useful where complex conditional distributions are suspected.

**Decision-relevant deterioration modelling, identifiability, and monitoring-system design.** Even when suitable deterioration and structural behavior models are available, the usefulness of monitoring-based inference depends on whether the monitored response contains sufficient information about the deterioration quantities relevant to the intended engineering decisions. The deterioration model should therefore represent the mechanisms and components that materially influence structural performance or lifetime-management decisions, but not necessarily at the finest physically conceivable resolution. If more independently uncertain parameters or deterioration states are introduced than can be supported by the available monitoring information, the resulting inference problem may become weakly identifiable and susceptible to overparameterization. Where physically justified, parameter pooling, hierarchical formulations, grouping of similar deterioration processes, sensitivity-based model reduction, or restriction of the candidate deterioration-state space can improve identifiability and computational robustness [6]. Such reductions should, however, be guided by physical knowledge and the decision problem rather than by computational convenience alone.

The required resolution of the deterioration model is consequently linked directly to the information provided by the monitoring system. The selected response quantities must be sufficiently sensitive to the deterioration states that need to be distinguished, while sensor type, number, placement, sampling strategy, measurement quality, and environmental or operational variability determine how reliably these response changes can be observed [5]. The deterioration model and monitoring system should therefore not be designed independently: the anticipated deterioration mechanisms and decision-relevant states should guide the selection of informative response quantities and sensor configurations, while limitations of the available monitoring information may motivate refinement or reduction of the deterioration representation. Decision-oriented approaches based on information gain or value of information provide a natural framework for this joint design problem [46, 47].

The numerical example illustrates this interaction particularly clearly. The deterioration model contains local fatigue processes at 22 hotspots, whereas the monitoring information consists of global modal properties. Under the adopted structural model, subcritical cracks do not affect the global stiffness and are therefore not directly observable from the modal data. Repeated observations corresponding to the intact structure nevertheless provide survival-type information by excluding or strongly downweighting deterioration trajectories that would have caused a brace failure before the current monitoring time. Once a change in modal properties is observed, several different local deterioration histories can remain compatible with essentially the same global structural response. The resulting ambiguity is therefore a genuine consequence of limited observability and identifiability and does not require the posterior distribution to separate into distinct modes. Moreover, dependence between the local deterioration processes allows evidence of deterioration at one location to modify the prognosis at other locations, as illustrated by the hotspot predictions following the year-20 observation. Where distinctions between the alternative deterioration states are decision-relevant, global response monitoring may need to be complemented by localized monitoring or inspection methods capable of resolving the remaining ambiguity or characterizing subcritical deterioration.

**Engineering significance, scope, and outlook.** The numerical example demonstrates how monitoring information can be propagated from parameter inference to quantities of direct engineering relevance. The posterior deterioration model provides the probabilistic basis for system-level diagnosis, crack-growth prognosis, hotspot failure probabilities, and structural reliability predictions within the same framework. In principle, uncertainty and ambiguity between alternative deterioration scenarios are retained throughout this propagation rather than being collapsed into a single inferred state.

The example also illustrates that uncertainty at the local deterioration level need not translate directly into the same form of uncertainty at the structural-system level. Several local deterioration histories may map to the same brace or system state, such that some local ambiguities disappear under the system-state mapping. Conversely, ambiguity between different brace configurations can remain relevant for subsequent prognosis and reliability prediction and must therefore be propagated when their relative probabilities can be represented reliably.

The distinction between local deterioration and structural consequence is also important from an engineering perspective. Failure of an individual hotspot or the associated loss of a brace does not necessarily imply system failure in a redundant structure. The structural significance of an inferred deterioration state must therefore be evaluated through the performance model, which maps the monitoring-informed deterioration state to structural resistance and failure probability. This separation enables the framework to distinguish between deterioration that is detectable or probable and deterioration that is critical for structural safety, thereby providing a more appropriate basis for risk-informed inspection, maintenance, or intervention decisions.

The present application is intentionally based on a controlled synthetic example and should therefore be interpreted as a demonstration of the probabilistic and computational framework rather than as validation of predictive accuracy for a field structure. Further work should address experimental and field validation, the integration of complementary inspection or local monitoring information, and the use of the resulting monitoring-informed predictions in inspection and maintenance planning. On the computational side, further work may investigate conditional-sampling strategies that combine efficient local sampling with broader exploratory transitions when the intermediate conditional distributions exhibit separated or otherwise complex regions [44]. Combining continuous global monitoring with targeted inspection could further allow inspection activities to be adapted according to the evolving probability of decision-relevant deterioration and its consequences, providing a natural link to value-of-information and lifetime-management approaches.

# 6 Conclusion

This paper presented a probabilistic framework for sequential diagnosis and prognosis of structural deterioration from response-based monitoring data. The framework separates inference of the latent deterioration process from the subsequent propagation of the inferred deterioration model to diagnosis, prognosis, and structural reliability prediction. For sequential Bayesian inference, BUS was extended through a nested formulation of the observation events associated with successively growing monitoring datasets, enabling Subset Simulation to continue from the conditional sample population obtained at the preceding monitoring stage. The numerical example demonstrated how monitoring information can be propagated from deterioration-parameter inference to system-level diagnosis, deterioration prognosis, and monitoring-informed reliability prediction while retaining uncertainty and ambiguity in the underlying deterioration state. Efficient Sequential BUS implementation depends on suitably chosen likelihood multipliers and adequate exploration of the intermediate conditional distributions. In the present application, tight stagewise likelihood multipliers can be determined directly from the discrete deterioration states; for more general problems, analytical or adaptive bounds and conditional samplers with enhanced global exploration provide natural directions for further development.

# Acknowledgements

I thank my colleagues Lukas Eichner and Patrick Simon from Division 7.2, Buildings and Structures, at the Bundesanstalt für Materialforschung und -prüfung (BAM) for their valuable support in implementing the numerical example, particularly the structural dynamic modelling and operational modal analysis.

## Declaration of generative AI use

I developed the modelling framework and algorithms presented in this work by integrating and extending established concepts and methods from structural health monitoring, deterioration modelling, Bayesian inference, and structural reliability. I used OpenAI's ChatGPT as an auxiliary tool for language editing, improving consistency of notation and terminology, and supporting interactive discussion and critical examination of technical formulations, interpretations, and their presentation. The scientific concepts, methodology, implementation, analyses, and conclusions are my own. I critically reviewed and independently verified all AI-assisted content and take full responsibility for the content of this article.

## Appendix A Model parameters and computational settings

The probabilistic model specifications and numerical parameter values used in the illustrative example are summarized in Table A.1. The computational settings used for Subset Simulation in both sequential inference and the subsequent monitoring-informed reliability analysis are given in Table A.2. Conditional sampling within Subset Simulation is performed in standard normal space using the adaptive conditional sampling algorithm proposed by Papaioannou et al. [45].

**Table A.1:** Probabilistic model specifications and numerical parameter values used in the illustrative example.

| Parameter | | Distribution | Values | |
|---|---|---|---|---|
| Annual maximum load [kN] | $S_{\max}$ | Gumbel | $\mu$ = 72.55 | COV = 0.35 |
| Design fatigue life [yr] | $T_{FL,j}$ | Deterministic | *see Figure 1* | |
| Partial safety factor for fatigue loads [-] | $\gamma_f$ | Deterministic | 1.00 | |
| Partial safety factor for fatigue strength [-] | $\gamma_m$ | Deterministic | 1.15 | |
| Fatigue limit (SN model) [-] | $\Delta$ | Normal | $\mu$ = 1.00 | $\sigma$ = 0.30 |
| Annual cycle rate [1/yr] | $\nu$ | Deterministic | $10^7$ | |
| Hotspot stress range (fatigue load) [N/mm$^2$] | $\Delta S_j$ | Weibull | $k$ = *calibrated* | $\lambda$ = 0.80 |
| Intercept of the first slope of the bilinear SN-curve with the log $N$-axis [corresponds to N/mm$^2$] | $\ln A$ | Normal | $\mu$ = 28.1277 | $\sigma$ = 0.4605 |
| Reference strength of SN curve [N/mm$^2$] | $\Delta S_c$ | Deterministic | 71 | |
| First negative inverse slope of bilinear SN curve [-] | $m_1$ | Deterministic | 3 | |
| Second negative inverse slope of bilinear SN curve [-] | $m_2$ | Deterministic | 5 | |
| Reference cycles of SN curve [-] | $N_c$ | Deterministic | $2 \cdot 10^6$ | |
| Reference cycles (at slope change) [-] | $N_q$ | Deterministic | $5 \cdot 10^6$ | |
| Paris' Law material parameter [corresponds to N/mm$^2$] | $\ln C_j$ | Normal | $\mu$ = *calibrated* | $\sigma$ = 0.514 |
| Paris' Law material parameter [-] | $m_j$ | Deterministic | *calibrated* | |
| Geometry factor in $SIF$ solution [-] | $Y_j$ | Deterministic | *calibrated* | |
| Initial crack depth [mm] | $A_{0,j}$ | Exponential | $\mu$ = 0.11 | |

| Parameter | | Distribution | Values | |
|---|---|---|---|---|
| Critical crack depth [mm] | $a_c$ | Deterministic | 20 | |
| Model uncertainty (fatigue load estimation) [-] | $B_{S,j}$ | Lognormal | $\mu = 1.00$ | $\sigma = 0.20$ |
| Model uncertainty ($SIF$ estimation) [-] | $B_{SIF,j}$ | Lognormal | $\mu = 1.00$ | $\sigma = 0.10$ |
| Common correlation coefficient for $A_{0,j}$ [-] | $\rho_{A_0}$ | Deterministic | 0.50 | |
| Common correlation coefficient for $B_{S,j}$ [-] | $\rho_{B_S}$ | Deterministic | 0.50 | |
| Common correlation coefficient for $B_{SIF,j}$ [-] | $\rho_{B_{SIF}}$ | Deterministic | 0.50 | |
| Common correlation coefficient for $\ln C_j$ [-] | $\rho_{\ln C}$ | Deterministic | 0.50 | |
| COV reflecting the uncertainty in the eigenvalue prediction [-] | $c_{\lambda,r}$ | Deterministic | 0.10 | |
| COV reflecting the uncertainty in the mode shape prediction [-] | $c_{\Phi,r}$ | Deterministic | 0.10 | |

**Table A.2:** Computational settings of Subset Simulation used for sequential inference and monitoring-informed reliability analysis in the illustrative example.

| Parameter | | Values |
|---|---|---|
| Number of samples per subset level | $N$ | 10,000 |
| Target conditional probability of intermediate events | $p_0$ | 0.1 |
| Number of seeds (Markov chains) | $N_s$ | 1000 |
| Chain length | $N_c$ | 10 |
| Number of samples between successive proposal adaptations | $N_a$ | 1000 |
| Initial standard deviation of the proposal distribution | $\sigma_0$ | 1.0 |
| Target acceptance rate | $a^*$ | 0.44 |
| Initial scaling parameter | $\lambda_0$ | 0.6 |

# References

[1] A. Rytter, "Vibrational Based Inspection of Civil Engineering Structures," PhD thesis, Department of Building Technology and Structural Engineering, Aalborg University, Denmark, 1993.

[2] S. Doebling, C. Farrar, and M. Prime, "A Summary Review of Vibration-Based Damage Identification Methods," *The Shock and Vibration Digest,* vol. 30, pp. 91–105, 03/01 1998, doi: 10.1177/058310249803000201.

[3] C. Farrar and K. Worden, *Structural Health Monitoring: A Machine Learning Perspective*. 2013.

[4] E. Viefhues, M. Döhler, F. Hille, and L. Mevel, "Statistical subspace-based damage detection with estimated reference," *Mechanical Systems and Signal Processing,* vol. 164, p. 108241, 2022/02/01/ 2022, doi: https://doi.org/10.1016/j.ymssp.2021.108241.

[5] P. Simon, R. Schneider, M. Baeßler, and G. Morgenthal, "A Bayesian Probabilistic Framework for Building Models for Structural Health Monitoring of Structures Subject to Environmental Variability," *Structural Control and Health Monitoring,* vol. 2024, no. 1, p. 4204316, 2024/01/01 2024, doi: https://doi.org/10.1155/2024/4204316.

[6] L. Eichner, R. Schneider, M. Baessler, and D. Kadoke, "Bayesian System and Damage Identification for an Offshore Jacket-Type Test Structure Using Heterogeneous SHM Data," *Struct Health Monit,* under review.

[7] A. Kamariotis *et al.*, "Monitoring-supported value generation for managing structures and infrastructure systems," *Data-Centric Engineering,* vol. 5, p. e27, 2024, Art no. e27, doi: 10.1017/dce.2024.24.

[8] S. Zorzi, M. Broccardo, D. Tonelli, and D. Zonta, "Reliability-based metrics for structural health monitoring information quality assessment," *Struct Health Monit,* vol. 24, no. 5, pp. 3028–3045, 2025/09/01 2025, doi: 10.1177/14759217241265956.

[9] W. H. Tang, "Probabilistic updating of flaw information," *Journal of Testing and Evaluation,* vol. 1, no. 6, pp. 459–467, 1973.

[10] H. O. Madsen, "Model updating in reliability theory," presented at the 5th International Conference on Applications of Statistics and Probability in Civil Engineering (ICASP 5), Vancouver, Canada, 1987.

[11] H. P. Hong, "Reliability analysis with nondestructive inspection," *Structural Safety,* vol. 19, no. 4, pp. 383–395, 1997.

[12] D. Straub, "Reliability updating with equality information," *Probabilistic Engineering Mechanics,* vol. 26, no. 2, pp. 254–258, 2011, doi: https://doi.org/10.1016/j.probengmech.2010.08.003.

[13] J. Luque and D. Straub, "Reliability analysis and updating of deteriorating systems with dynamic Bayesian networks," *Structural Safety,* vol. 62, pp. 34–46, 2016, doi: http://dx.doi.org/10.1016/j.strusafe.2016.03.004.

[14] R. Schneider, S. Thöns, and D. Straub, "Reliability analysis and updating of deteriorating systems with subset simulation," *Structural Safety,* vol. 64, pp. 20–36, 2017, doi: http://dx.doi.org/10.1016/j.strusafe.2016.09.002.

[15] J. L. Beck and L. S. Katafygiotis, "Updating Models and Their Uncertainties. I: Bayesian Statistical Framework," *Journal of Engineering Mechanics,* vol. 124, no. 4, pp. 455–461, 1998.

[16] L. S. Katafygiotis, C. Papadimitriou, and H.-F. Lam, "A probabilistic approach to structural model updating," *Soil Dynamics and Earthquake Engineering,* vol. 17, no. 7, pp. 495–507, 1998/10/01/ 1998, doi: https://doi.org/10.1016/S0267-7261(98)00008-6.

[17] J. L. Beck and S. K. Au, "Bayesian updating of structural models and reliability using Markov chain Monte Carlo simulation," *J Eng Mech-Asce,* vol. 128, no. 4, pp. 380–391, 2002, doi: 10.1061/(Asce)0733-93399.

[18] M. W. Vanik, J. L. Beck, and S. K. Au, "Bayesian probabilistic approach to structural health monitoring," (in English), *Journal of Engineering Mechanics,* vol. 126, no. 7, pp. 738–745, Jul 2000, doi: Doi 10.1061/(Asce)0733-9399(2000)126:7(738).

[19] K.-V. Yuen, J. L. Beck, and S. K. Au, "Structural damage detection and assessment by adaptive Markov chain Monte Carlo simulation," *Structural Control and Health Monitoring,* vol. 11, no. 4, pp. 327–347, 2004/10/01 2004, doi: https://doi.org/10.1002/stc.47.

[20] I. Behmanesh, B. Moaveni, G. Lombaert, and C. Papadimitriou, "Hierarchical Bayesian model updating for structural identification," *Mechanical Systems and Signal Processing,* vol. 64-65, pp. 360–376, 2015/12/01/ 2015, doi: https://doi.org/10.1016/j.ymssp.2015.03.026.

[21] A. Kamariotis, E. Chatzi, and D. Straub, "Value of information from vibration-based structural health monitoring extracted via Bayesian model updating," *Mechanical Systems and Signal Processing,* vol. 166, p. 108465, 2022/03/01/ 2022, doi: https://doi.org/10.1016/j.ymssp.2021.108465.

[22] A. Kamariotis, E. Chatzi, and D. Straub, "A framework for quantifying the value of vibration-based structural health monitoring," *Mechanical Systems and Signal Processing,* vol. 184, p. 109708, 2023/02/01/ 2023, doi: https://doi.org/10.1016/j.ymssp.2022.109708.

[23] A. Kamariotis, L. Sardi, I. Papaioannou, E. Chatzi, and D. Straub, "On off-line and on-line Bayesian filtering for uncertainty quantification of structural deterioration," *Data-Centric Engineering,* vol. 4, p. e17, 2023, Art no. e17, doi: 10.1017/dce.2023.13.

[24] D. Straub and I. Papaioannou, "Bayesian updating with structural reliability methods," *Journal of Engineering Mechanics,* vol. 141, no. 3, 2015. [Online]. Available: http://dx.doi.org/10.1061/(ASCE)EM.1943-7889.0000839.

[25] S.-K. Au and J. L. Beck, "Estimation of small failure probabilities in high dimensions by subset simulation," *Probabilistic Engineering Mechanics,* vol. 16, no. 4, pp. 263–277, 2001.

[26] D. Straub, I. Papaioannou, and W. Betz, "Bayesian analysis of rare events," *Journal of Computational Physics,* vol. 314, pp. 538–556, 2016, doi: http://dx.doi.org/10.1016/j.jcp.2016.03.018.

[27] R. Schneider, *Time-variant reliability of deteriorating structural systems conditional on inspection and monitoring data* (BAM-Dissertationsreihe, no. 168). Berlin: Bundesanstalt für Materialforschung und -prüfung (BAM), 2020, pp. 1–188.

[28] D. J. Jerez, H. A. Jensen, C. Figueroa, and J. Chen, "Reliability updating of structural dynamical systems under stochastic excitation using a two-phase subset simulation approach," *Mechanical Systems and Signal Processing,* vol. 250, p. 114165, 2026/04/15/ 2026, doi: https://doi.org/10.1016/j.ymssp.2026.114165.

[29] E. Simoen, C. Papadimitriou, and G. Lombaert, "On prediction error correlation in Bayesian model updating," *Journal of Sound and Vibration,* vol. 332, no. 18, pp. 4136–4152, 2013/09/02/ 2013, doi: https://doi.org/10.1016/j.jsv.2013.03.019.

[30] I. Koune, Á. Rózsás, A. Slobbe, and A. Cicirello, "Bayesian system identification for structures considering spatial and temporal correlation," *Data-Centric Engineering,* vol. 4, p. e22, 2023, Art no. e22, doi: 10.1017/dce.2023.18.

[31] D. Straub, R. Schneider, E. Bismut, and H.-J. Kim, "Reliability analysis of deteriorating structural systems," *Structural Safety,* vol. 82, p. 101877, 2020, doi: https://doi.org/10.1016/j.strusafe.2019.101877.

[32] V. A. Zayas, S. A. Mahin, and E. P. Popov, "Cyclic inelastic behavior of steel offshore structures," University of California, Berkley, USA, 1980.

[33] T. H. Søreide, J. Amdahl, E. Eberg, T. Hovland, and Ø. Hellan, "USFOS - A Computer Program for Progressive Collapse Analysis of Steel Offshore Structures: Theory Manual," SINTEF, Trondheim, Norway, 1988.

[34] M. Amiri and B. Namdar Zangeneh, „ "Structural damage detection using modal strain energy," presented at the 8th International Conference on Coasts, Ports and Marine Structures (ICOPMAS 2008), Tehran, Iran, 24–26 November 2008, 2008.

[35] E. Bismut, J. Luque, and D. Straub, "Optimal priorization of inspections in structural systems considering component interactions and interdependence," presented at the 12th International Conference on Structural Safety & Reliability (ICOSSAR 2017), Vienna, Austria, 2017.

[36] D. Straub, "Generic Approaches to Risk Based Inspection Planning for Steel Structures," Dr. sc. techn. PhD thesis, ETH Zürich, Switzerland, 2004.

[37] F. McKenna, "OpenSees: A Framework for Earthquake Engineering Simulation," *Computing in Science & Engineering,* vol. 13, no. 4, pp. 58–66, 2011, doi: 10.1109/MCSE.2011.66.

[38] M. Zhu, F. McKenna, and M. H. Scott, "OpenSeesPy: Python library for the OpenSees finite element framework," *SoftwareX,* vol. 7, pp. 6–11, 2018/01/01/ 2018, doi: https://doi.org/10.1016/j.softx.2017.10.009.

[39] W. Betz, I. Papaioannou, J. L. Beck, and D. Straub, "Bayesian inference with Subset Simulation: Strategies and improvements," *Computer Methods in Applied Mechanics and Engineering,* vol. 331, pp. 72–93, 4/1/ 2018, doi: https://doi.org/10.1016/j.cma.2017.11.021.

[40] B. Peeters and G. De Roeck, "Stochastic System Identification for Operational Modal Analysis: A Review," *Journal of Dynamic Systems, Measurement, and Control,* vol. 123, no. 4, pp. 659–667, 2001, doi: 10.1115/1.1410370.

[41] D. Straub, "Reliability assessment of deteriorating structures: challenges and (some) solutions," presented at the 6th International Symposium on Life-Cycle Civil Engineering (IALCCE 2018), Ghent, Belgium, 2018.

[42] A. Artinov *et al.*, "Hydrogen-Induced Stress Corrosion Cracking in Quenched and Tempered Prestressing Steel Wires: Fatigue Crack Growth and Fracture-Mechanics-Based Residual Life Assessment," in *Physics-informed Structural Health Monitoring - Proceedings of CSHM-10 Workshop*, vol. XXX, M. Baeßler, R. Herrmann, F. Hille, E. Niederleithinger, and J. F. Unger Eds., (Lecture Notes in Civil Engineering. Berlin, Germany: Springer Nature Switzerland, 2026.

[43] K. Breitung, "The geometry of limit state function graphs and subset simulation: Counterexamples," *Reliability Engineering & System Safety,* vol. 182, pp. 98–106, 2019, doi: https://doi.org/10.1016/j.ress.2018.10.008.

[44] A. Sharma, P. Chakroborty, and M. D. Shields, "SuSIE: Subset Simulation with Intrepid Exploration," *arXiv,* vol. arXiv:2607.22989v1, 2026/07/28 2026. [Online]. Available: https://arxiv.org/abs/2607.22989.

[45] I. Papaioannou, W. Betz, K. Zwirgelmaier, and D. Straub, "MCMC Algorithms for Subset Simulation," *Probabilistic Engineering Mechanics,* vol. 41, pp. 89–103, 2015.

[46] L. Eichner, R. Schneider, and M. Baeßler, "Optimal vibration sensor placement for jacket support structures of offshore wind turbines based on value of information analysis," *Ocean Engineering,* vol. 288, p. 115407, 2023.

[47] N. E. Silionis and K. N. Anyfantis, "On decision-theoretic model assessment for structural deterioration monitoring," *Mechanical Systems and Signal Processing,* vol. 222, p. 111776, 2025/01/01/ 2025, doi: https://doi.org/10.1016/j.ymssp.2024.111776.